\documentclass[a4paper,fleqn]{cas-dc}

\usepackage{savesym}
\savesymbol{comment}

\usepackage[markup=default,authormarkup=none]{changes}

\restoresymbol{CHANGES}{comment}

\usepackage[numbers]{natbib}

\usepackage{cleveref}
\usepackage{siunitx}
\usepackage[all]{nowidow}
\usepackage{tabularx}
\usepackage{booktabs}
\usepackage[title]{appendix} 

\input{aas_macros.sty}

\renewcommand{\vec}[1]{\mbox{\boldmath$#1$}}

\def \d			{\mathop{}\!\mathrm{d}}							

\DeclareSIUnit \arcmin 	{arcmin}
\DeclareSIUnit \arcsec 	{arcsec}
\DeclareSIUnit \parsec 	{pc}
\DeclareSIUnit \eV 			{eV}
\DeclareSIUnit \Msun 		{M_\odot}
\DeclareSIUnit \cts			{cts}
\DeclareSIUnit \deg			{deg}
\DeclareSIUnit \ph 			{ph}

\definechangesauthor[name={Moline}, color=red]{am}
\definechangesauthor[name={cyan}, color=red]{cy}
\definechangesauthor[name={Yunis}, color=red]{ry}
\definechangesauthor[name={Krut}, color=red]{ak}
\definechangesauthor[name={Arguelles}, color=red]{ca}
\definechangesauthor[name={Rueda}, color=red]{jar}
\definechangesauthor[name={Mavromatos}, color=red]{nm}

\begin{document}
\let\WriteBookmarks\relax
\def\floatpagepagefraction{1}
\def\textpagefraction{.001}
\shorttitle{\boldmath Galactic Center constraints on self-interacting sterile neutrinos from fermionic dark matter (``ino'') models}
\shortauthors{R. Yunis et~al.}

\title[mode=title]{\boldmath Galactic Center constraints on self-interacting sterile neutrinos from fermionic dark matter (``ino'') models}

\author[a,b]{R. Yunis}
\ead{yunis121@gmail.com}

\author[c]{C.~R. Arg\"uelles}

\author[d]{N.~E. Mavromatos}

\author[e,f,g]{\'A. Molin\'e}

\author[a]{A. Krut}

\author[a,b]{M. Carinci}

\author[a,b,h,i,j]{J.~A. Rueda}

\author[a,b,k]{R. Ruffini}

\address[a]{ICRANet, Piazza della Repubblica 10, I--65122 Pescara, Italy}
\address[b]{ICRA, Dipartimento di Fisica, Sapienza Università di Roma, P.le Aldo Moro 5, I--00185 Rome, Italy}

\address[c]{Facultad de Ciencias Astron\'omicas y Geof\'isicas, Universidad Nacional de La Plata, Paseo del Bosque, B1900FWA La Plata, Argentina}

\address[d]{Theoretical Particle Physics and Cosmology Group, Department of Physics, King's College London, Strand WC2R 2LS, London, U.K.}

\address[e]{Instituto de Física Teórica UAM-CSIC, Universidad Autónoma de Madrid, C/ Nicolás Cabrera, 13-15, 28049 Madrid, Spain}
\address[f]{Departamento de Física Teórica, M-15,Universidad Autónoma de Madrid, E-28049 Madrid, Spain}

\address[g]{Instituto de Astrofísica e Ciências do Espaço, Faculdade de Ciências da Universidade de Lisboa, Edifício C8, Campo Grande, P-1749-016 Lisbon, Portugal}

\address[h]{ICRANet-Ferrara, Dipartimento di Fisica e Scienze della Terra, Universit\`a degli Studi di Ferrara, Via Saragat 1, I--44122 Ferrara, Italy}
 
\address[i]{Dipartimento di Fisica e Scienze della Terra, Universit\`a degli Studi di Ferrara, Via Saragat 1, I--44122 Ferrara, Italy}

\address[j]{INAF, Istituto de Astrofisica e Planetologia Spaziali, Via Fosso del Cavaliere 100, 00133 Rome, Italy}

\address[k]{INAF, Viale del Parco Mellini 84, 00136 Rome  Italy}


\begin{abstract}
 The neutrino minimal standard model ($\nu$MSM) has been tightly constrained in the recent years, either from dark matter (DM) production or from X-ray and small-scale observations. However, current bounds on sterile neutrino DM can be significantly modified when considering a $\nu$MSM extension, in which the DM candidates interact via a massive (axial) vector field. In particular, standard production mechanisms in the early Universe can be affected through the decay of such a massive mediator.
We perform an indirect detection analysis to study how the $\nu$MSM parameter-space constraints are affected by said interactions. We compute the X-ray fluxes considering a DM profile that self-consistently accounts for the particle physics model by using an updated version of the Ruffini-Argüelles-Rueda (RAR) fermionic (“ino”) model, instead of phenomenological profiles such as the Navarro-Frenk-White (NFW) distribution. 
We show that the RAR profile accounting for interacting DM, is compatible with measurements of the Galaxy rotation curve and constraints on the DM self-interacting cross section from the Bullet cluster. A new analysis of the X-ray NuSTAR data in the central parsec of the Milky Way, is here performed to derive constraints on the self-interacting sterile neutrino parameter-space. Such constraints are stronger than those obtained with commonly used DM profiles, due to the dense DM core characteristic of the RAR profiles.
\end{abstract}

\maketitle


\section{Introduction}
\label{sec:intro}

While the evidence for the existence of dark matter (DM) is implied from astrophysical and cosmological observations of gravitational effects, a huge effort is still focused on the understanding of the nature of the particles that make up this unknown matter as well as its detection~\cite{Jungman:1995df, Bergstrom:2000pn, Munoz:2003gx, Bertone:2004pz}. Among the myriad of DM candidates proposed, a sterile neutrino with a mass in the $\si{\kilo\eV}$ range has been claimed as a viable candidate, falling in the category of \textit{warm} DM (WDM)~(see~\cite{Berezhiani:1995, Langacker:1998, Asaka:2005, Aliu:2005}  for some relevant examples and \cite{2017JCAP...01..025A, 2012PDU.....1..136B} for a review).

These claims seem to be revitalized given recent results about neutrino oscillations and several other physics phenomenona which are not predicted by the Standard Model (SM) and suggest new, unknown physics~\cite{Ahmad:2002, Ashie:2005, Araki:2005, Weinheimer:2003}. In particular, a minimal extension of the SM, the so-called \textit{Neutrino Minimal Standard Model} ($\nu$MSM) introduces three sterile right handed neutrinos, where the lightest one might account for the presence of DM in the universe~\cite{Asaka:2005}. 

From the gravitational evidence about the existence of DM, we can infer that the DM particle  must be stable on cosmological time scales. Nevertheless, huge amounts of DM particles can decay even for such extremely long life-times and the decay signals
may be in the observable range to be detectable
(see e.g.
~\cite{Eichler:1989, 2009JCAP...01..043N, PerezPileviez, Ibarra:2010, Covi:2009, HISANO2009101, Buckley:2010, Demir:2010}). 
Within the framework of the $\nu$MSM, the hypothetical sterile neutrino is an example of decaying DM with a life-time several times greater than the age of the Universe~\cite{Asaka:2005}. 
Viability for such sterile neutrinos as to constitute the entirety of cosmological DM requires low enough mixing with the standard sector, measured by the \textit{mixing angle}

\begin{equation}
	\label{mixingangle}
	\theta^2=\sum_{\alpha=e,\mu,\tau}(v^2 F_{\alpha,1})/m_{s}^2\, ,
\end{equation}
with $v^2$ the Higgs boson v.e.v., $F_{\alpha,1}$ the Yukawa couplings for the right-chiral neutrinos and $m_{s}$ the sterile neutrino mass. Indeed, it is enough for $\theta^2$ to fall below $\num{2.5E-13} (\si{\mega\eV}/m_{s})^5$ as to have a life-time longer than the age of the universe for tree-level decays~\cite{Dolgov02}. More stringent observational bounds, as the ones arising from the diffuse X/$\gamma$-ray background, ensure the fulfillment of this life-time condition~\cite{Dolgov02}.
These bounds are significantly overcome by the ones due to sterile neutrino production mechanisms (not displayed in \cref{fig:RAR_vs_P2017}).
Both the sterile neutrino mass $m_s$, and the mixing angle between active and sterile neutrinos $\theta$, constitute the parameter space for $\nu$MSM  regarding DM. Their decay open the possibility of indirectly detecting DM via the identification of such interactions products as photons or neutrinos. Indeed, the sterile neutrinos would have a subdominant radiative decay channel into a photon and a light (mostly active) neutrino~\cite{Pal:1982,BARGER1995365}. An important clue for searching for the decay of a sterile neutrino candidate may be coming from the X-ray observation of DM-dominated objects, such as galaxies and clusters of galaxies. The region of the galaxy with the highest DM density is the Milky Way center and constitutes the classical target for DM searches
, though it is often made more difficult by the complicated astrophysics involved
 (see e.g.~\cite{2010pdmo.book.....B} for a compilation of works). The Galactic center (GC) region has been extensively studied in the realm of sterile neutrino DM decays using observations of several X-ray satellites such as Suzaku, Chandra, XMM-Newton (see \cite{2017JCAP...01..025A} for a recent review), as well as from the NuSTAR mission \cite{Perez2017}. 

The above-cited works aimed at constraining the $\nu$MSM parameter space using X-ray observations from both galactic and extragalactic objects. Such limits come as complementary to the ones imposed by production mechanisms of DM in the early universe, such as non-resonant (Dodelson - Widrow) production \cite{DodelsonWidrow94,AsakaLaineShaposhnikov,2017JCAP...01..025A} and resonant production \cite{LaineShaposhnikov,ShiFuller,AsakaLaineShaposhnikov}. Other limits to the $\nu$MSM model include phase space distribution bounds, as well as bounds from local group galaxy counts, which exclude masses below several $\si{\kilo\eV}$ \cite{BoyarskyLowerBound,HoriuchiHumphrey,Schneider}. 

X-ray bounds provide upper limits to the mixing angle $\theta$ (between dark and active sector) as a function of particle mass $m_{s}$. 
The pertinent constraints are summarized in figure 1 in \cite{Perez2017}: upper bounds to the mixing angle $\theta$ can be set using X-ray searches, while lower bounds are placed by ensuring correct sterile neutrino abundances to account for the whole DM budget. Also, phase space constraints and MW satellite counts can place lower bounds in the sterile neutrino mass.


We observe from that figure that the allowed regime of masses for the lightest sterile neutrino within the framework of the conventional $\nu$MSM~\cite{nuMSM} are in the range 
\begin{equation}\label{constrnmsm}
10~{\rm keV} \,  \lesssim \,  m_s \,  \lesssim \, 16~{\rm keV}~. 
\end{equation}
Such a narrow range seems to marginally include the tentantive $3.5$~keV signal for DM (corresponding to $m_s \sim 7 ~{\rm keV}$~\cite{Bulbul2014,Boyarsky:2014jta}).


In this paper, we focus on placing constraints using these X-ray signals due to sterile neutrino decays within an extension of the $\nu$MSM framework, in which the DM candidates interact via a dark-sector massive (axial) vector field. Such an extension, proposed in \cite{amrr} by Arg\"uelles-Mavromatos-Ruffini-Rueda (AMRR from now on), represents a minimal extension to the $\nu$MSM model and involves the assumption of novel DM density profiles which depend on the particle mass and the interaction coupling constant (as discussed in \cref{sec:interactingDM}). The relevance and main motivation of adding self interactions to the model, is because it significantly modifies the production mechanisms of sterile neutrinos, relaxing current $\nu$MSM bounds as discussed in \cref{sec:interactingDM} (see \cite{deGouvea2019} for an analogous result). Besides, it can affect the shape of the DM density profiles, and alter the cosmological evolution with its corresponding imprints on structure formation.

A crucial point is that the intensity of the DM decay flux expected from an individual halo depends mainly on the DM density distribution inside it. N-body simulations within the $\Lambda$CDM paradigm, seem to point towards a single universal description for the DM halo density profiles, and different parameterizations had been obtained in the literature~\cite{Navarro:1995iw,Navarro:1996gj,MNR:MNR3039,Klypin:2000hk,Navarro:2003ew}. However, in this work we are interested in particles pertaining to the WDM paradigm that decouple while still relativistic, implying a subsequent free-streaming which damps primordial density fluctuations below a cutoff scale sensitive to the particle mass (see e.g.~\cite{Boyarsky2009}). For \si{\kilo\eV}-ish particles, such a damping implies a suppression in the power spectrum which goes from $\sim \SI{10}{\percent}$ at \si{\mega\parsec} scales when compared with the $\Lambda$CDM one, and becomes stronger on scales below \SI{E2}{\kilo\parsec} (see~\cite{2012PDU.....1..136B} for current constraints). One of the main consequences of such a suppression is the difference in morphology of the DM density profiles: while CDM (on DM-only simulations) halos (and sub-halos) are cuspy through the center, the WDM ones tend to form cored inner halos for low enough particle masses \textit{below} \si{\kilo\eV} (see e.g.~\cite{2012MNRAS.424..684S}). Nevertheless, in the case of few to several \si{\kilo\eV}, recent high resolution N-body simulations \cite{2017MNRAS.464.4520B} developed to understand the small-scale structure differences between CDM and WDM cosmologies, show that the density profiles in WDM halos with masses $M\gtrsim \SI{E10}{\Msun}$\footnote{
	Less massive WDM halos do start to show systematically lower concentrations with respect to CDM ones \cite{2017MNRAS.464.4520B}.
} are indistinguishable from their CDM counterparts, and well fitted by Einasto profiles~\cite{Navarro:2003ew}, or even by NFW \cite{2012MNRAS.424..684S} (though for much larger halo masses in the latter). Therefore, in this paper we will compare  the Navarro-Frenk-White (NFW) profile~\cite{Navarro:1995iw, Navarro:1996gj}, the Einasto (EIN) profile~\cite{Einasto2006,Navarro:2003ew} and the Burkert (BUR) profile~\cite{Burkert:1995}, with the recently proposed Ruffini-Arg\"uelles-Rueda (RAR) profile~\cite{2015MNRAS.451..622R,Arguelles2016} in order to bracket the theoretical uncertainty in the limits from the modeling of the expected DM decay flux of sterile neutrino WDM.

At this point it is important to emphasize that the robustness of the constraints obtained here depends crucially on the precise knowledge of the DM profile at the GC.
In this line, we would like to point out some advantages of using the RAR model, over phenomenological DM profiles (e.g. NFW~\cite{Navarro:1995iw, Navarro:1996gj},  Einasto~\cite{Einasto2006}, etc), regarding the self-consistent estimation of the Galactic distribution of \si{\kilo\eV} WDM fermions (``inos'').

While phenomenological DM profiles arise from the fitting of (classical) N-body simulation with finite spatial resolution down to $\sim \SI{0.1}{\kilo\parsec}$ scales, when aiming to emission regions at parsec-scales, extrapolated versions of those profiles are used due to the lack of knowledge of the DM content (and precise distribution) around SgrA*. Instead, in the RAR model (and generalized versions) the Pauli principle is self-consistently included from the phase-space distribution function at relaxation, leading to a continuous \textit{dense core} -- \textit{diluted halo} profile from the very center (without the need of the BH) all the way to the outer halo. Such DM profiles are successfully applied to explain the rotation curves as well as different universal relations from dwarfs to ellipticals including the Milky Way, as shown in \cite{2015MNRAS.451..622R,Arguelles2016,Arguelles2018}. Moreover, at difference with the \textit{pseudo} particles involved in numerical simulations, the building blocks of the RAR profiles are fermionic particles which obey a well-established nature and equation of state. This allows us to make direct contact with particle physics scenarios, e.g. $\sim$ keV sterile neutrino physics (see discussion in \cite{amrr}).  

We would also like to mention some additional effects/ingredients that have been included in N-body simulations via the use of \textit{pseudo-particles}. There remains open the possibility of performing analogous simulations, introducing the more complicated quantum field-theoretical nature of the inos, either \textit{free} (original RAR~\cite{2015MNRAS.451..622R,Arguelles2016,Arguelles2018}), or \textit{self-interacting} (RAR+SIDM/AMRR~\cite{amrr}), to assess their possible effects on the halo formation and dynamics:

(i) Deviation either from spherical-symmetry and/or isotropy (in velocity) at virialization. Halo ellipticity and velocity anisotropy are not usually considered when obtaining sterile neutrino mass bounds, though in \cite{Boyarsky2008} the impact of typical ellipticity values \cite{Martin2008,vanUitert2016} has been evaluated within traditional profiles. In \cite{IngrossoMerafina92}, the effects of velocity anisotropy have been included in what we can call a Newtonian version of the RAR model, with marginal effects with respect to the isotropic results when assessed within the limits of observational uncertainties. This agrees with the fact that the RAR profiles provide very good fits for a large plethora of galactic and extragalactic data as shown in \cite{Arguelles2016,Arguelles2018}.

(ii)
From the cosmological point of view, the available amount of DM particles in the galaxies in modern eras of the Universe depends on detailed calculations of the entire cosmological history of microscopic models. Numerical simulations of DM halos formation including merging history physics have been performed both within CDM and WDM cosmologies (see, e.g., \cite{Ludlow2016}, and references therein for details and for a recent account). Analogous calculations/simulations for the RAR-model inos, accounting for their quantum nature and self-interactions, are starting to be developed by some of the authors in \cite{2020arXiv200205778Y}, though are well beyond the scope of the present work. 

While waiting for a full understanding of such cosmological simulations for quantum particles, we have an important hint from the fact that Fermi Dirac-like phase-space distributions (such as the one given by Eq.~\ref{eq:FDc-PS}, leading to the \textit{core -- halo} RAR profile), has been shown to be a possible outcome of the process of collissionless relaxation within cosmological timescales. This kind of distribution function can be a stationary solution of a generalized thermodynamic Fokker-Planck equation for fermions, including the physics of violent relaxation and evaporation \cite{1998MNRAS.300..981C,Chavanis2004}, appropriate within the non-linear structure formation process. Importantly, the equilibrium RAR \textit{core -- halo} profiles \cite{Arguelles2016} can be shown to be thermodynamically stable and extremely long-lived (i.e. maximising a coarse-grained entropy within cosmological timescales), as an outcome of violent relaxation processes, thus being a reachable end-state in Nature \cite{daky}. We refer the reader to \cref{sec:RAR} for further details on this topic.

It is then clear that the RAR model leads to a DM profile that presents distinct qualitative and quantitative features that allows model parameters to be constrained in a precise way. We mention for example the explicit particle mass dependence in the profile itself, and a novel \textit{dense quantum core - diluted halo} morphology which, for $mc^2 = \SIrange{48}{345}{\kilo\eV}$, can provide excellent fits to the Milky Way rotation curve data with the central DM core being an alternative to the BH scenario in SgrA*~\cite{Arguelles2016}.

As masses larger than \SI{48}{\kilo\eV} are above the upper particle mass limit set by the currently existing bounds on the $\nu$MSM parameter space \cite{Perez2017,2017JCAP...01..025A}, it is clear that 
\textbf{(A)} the microphysical model itself should be altered, or \textbf{(B)} there are more than one species of DM particles (for recent reviews see \cite{2017IJMPD..2630007M} and references therein), which would then imply a smaller portal mixing angle within the $\nu$MSM, thus relaxing the upper bound on masses, as obtained from x-ray constraints, or \textbf{(C)} a consistent extension to the RAR DM profile allowing for $m < \SI{48}{\kilo\eV}$ must be considered \footnote{Lower particle masses between $\sim 10$ and $48$~keV can fit as well the MW rotation curve\cite{Arguelles2016}, but the central core does not provide the BH alternative. However, a self-gravitating DM concentration around a BH is a dynamical (open) problem that requires considering accretion processes (e.g. solving relativistic Euler equations coupled to Einstein), which deserves a more dedicated analysis out of the scope of the present work.}.
In this work, we will study in detail the case
(A). We will discuss the constraints on the RAR profile coming from X-ray observations 
going beyond the $\nu$MSM scenario, by including self-interactions among the fermions in \cref{sec:interactingDM}, using the AMRR model~\cite{amrr}, which identifies the RAR DM fermions with the (self-interacting) sterile right-handed neutrinos. 
We will show that within the RAR+Self-interacting DM (RAR+SIDM/AMRR) approach, it is possible to be in agreement with the total MW rotation curve, while respecting the bullet cluster constraints.

We further estimate new upper bounds in the mixing angle $\theta$ for $m_s \geq 48$ keV, but in this case under the RAR+SIDM model assumption. These limits are obtained by comparing the photon flux observations from the GC against the (DM halo model dependent) theoretical expected one. In particular, we focus on searches with NuSTAR satellite which have provided accurate observation of diffuse X-ray emission within the few central parsecs around SgrA*~\cite{Mori2015}. A similar analysis from the NuSTAR data was used in \cite{Perez2017} in the $\nu$MSM standard scenario, though adopting corona-like regions around the GC, where diffuse photons are included but excluding the very center (where many individual bright X-ray sources can be identified).

Thus, in this work we aim to revisit X-ray constraints for decays of sterile neutrinos in the context of the AMRR extension to the $\nu$MSM model. This is motivated by the possibility of (i) relaxing existing constraints on the parameter space for $\nu$MSM (potentially allowing for higher particle masses or lower mixing angles); (ii) allowing us to intrinsically link the DM profile with the underlying particle physics model including DM self-interactions (thus requiring a fundamentally different indirect detection analysis due to the particle mass dependence in the profile). 

The outline of the work is as follows: in \cref{sec:decay} we describe the calculation of the expected DM decay flux. We summarize the relevant ingredients to compute it and the choice of parameters of the $\nu$MSM. We consider several parametrizations for the DM density profile and discuss its critical role in the expected DM decay flux.
In \cref{sec:interactingDM}, we discuss the effects in the relaxation of these bounds due to hidden dark sector interactions, proposed by AMRR in \cite{amrr} and extended here for the more realistic version of the RAR+SIDM model which includes escape of particle effects \cite{Arguelles2016}. 
Using the X-ray observations from the GC, in \cref{sec:bounds} we obtain the revised upper limits on the sterile neutrino mixing angles as a function of particle mass.
Finally, in \cref{sec:concls} we draw our conclusions.


\section{Sterile neutrino decay: X-ray flux}
\label{sec:decay}

The radiative sterile neutrino decay channel to a photon  and an active neutrino produces a spectral line in the X-ray. The decay width $\Gamma$, is defined as~\cite{Boyarsky2009,Pal:1982, BARGER1995365}
\begin{equation}
\Gamma = \frac{9}{1024}\frac{\alpha G_{\rm F}^{2}}{\pi^{4}}m_{s}^5\left | \Theta \right |^{2} ~
\label{eq:decay_rate}
\end{equation}
with $\alpha$  the fine-structure constant, $\left | \Theta \right |^{2} = \sin^2(2\theta)$ and $G_{\rm F}$ the Fermi constant.
Here $\theta$ is typically a small quantity below the electroweak scale, as arising from Cosmic X-ray Background (CXB) constraints ~\cite{Abazajian:2007}. The expected energy flux observed from the decay of a massive neutrino will depend on both the distance and distribution of DM across the field of view of the detector, as well as on the parameter space ($\theta$, $m_{s}$).

This decay channel is due to mixing between active and sterile sectors under the $\nu$MSM model. The interaction arises due to mass mixing thanks to the addition of a Majorana mass term for the sterile neutrinos, plus a Yukawa Higgs-portal term as:
\begin{equation}\label{yuk_s2}
{\mathcal L}_{\rm Yuk} = F_{\alpha I} \, {\bar \ell}_\alpha \, N_{R\, I} \phi^c   + {\rm h.c.}~,  \quad I=1,2,3
\end{equation}
where $\ell_\alpha$ are the lepton doublets of the Standard Model (SM), $\alpha = e, \mu, \tau$, $F_{\alpha I}$ are the appropriate Yukawa couplings which relates to the mixing angle $\theta$ via \cref{mixingangle} above, and $\phi^c$ is the SM conjugate Higgs field, \emph{i.e}. $\phi^c = i \tau_2 \phi^\star$ (with $\tau_2 $ the $2 \times 2$ Pauli matrix), with $N_{R\, I}$ the three sterile neutrino fields. Further details will be explored in \cref{sec:interactingDM}.

If $x$ denotes the linear coordinate along the line of sight (l.o.s.) and $d\Omega$ the solid angle element of the detector field of view, the differential photon flux from a volume element in the galaxy $x^{2} dx d\Omega$,  reaching a unit effective area of the detector is proportional to the DM density profile $\rho$, and given by \footnote{
	Here, we have ignored general relativistic effects on proper volume integrals, as well as a factor of $dN/dE$ given that the emission is almost a single line (compared to the resolution of the detectors, see the discussion in \cite{Perez2017})
} 
\begin{equation}
\d f=x^{2}\d x \d\Omega\frac{\Gamma }{m_{s}}\frac{\rho(r)}{4\pi x^2} ~,
\end{equation}
since each volume element contains $\rho(r)/m_{s}$  sterile neutrinos. The average flux observed in a solid angle is then found by integrating over the $\rho$  along the line of sight connecting the detector and the GC and the solid angle,
\begin{equation}
F=\frac{\Gamma}{4 \pi m_{s}}\int_{\Omega_{\rm l.o.s.}}  \d\Omega \int \rho (r(x,\Omega))\, \d x ~.
\label{eq:exp_flux}
\end{equation}
This expression can be cast as
\begin{equation}
F=\frac{\Gamma}{4 \pi m_{s}} S_{\rm DM} ~,
\label{eq:S_factor_definition}
\end{equation}
where the $S_{\rm DM}$ factor contains both the features of the DM profile of interest and the observation details such as the location of the observed region itself. The remaining factor depends exclusively on the particle physics parameters and $\nu$MSM specific decay rates. By asking $F^{\rm obs}_{\rm max}\geq F$ with $F^{\rm obs}_{\rm max}$ the maximum observed X-ray flux (see \cref{sec:signal}), it is possible to place upper limits on the sterile neutrino mixing angle as a function of particle mass, as shown in \cref{fig:RAR_vs_P2017}. In the following sections, we will discuss the ingredients required to place these bounds: we will discuss several models for DM distribution profiles in sections \ref{sec:profiles} and \ref{sec:RAR}, as well as the AMRR extension in section \ref{sec:interactingDM}, assessing their impact on $S_{\rm DM}$ (see also appendix \ref{appendix:SDM} for details). Finally, in section \ref{sec:signal} we calculate the maximum observed X-ray flux $F^{\rm obs}_{\rm max}$ for different observation targets around the Galaxy center.

\subsection{\texorpdfstring{$\nu$}{nu}MSM parameter bounds}
\label{sec:comparison}

All the constraints calculated in this work are upper limits on the mixing angle as a function of the particle mass, but several other parameter-space bounds exist, under the particle model here considered. Phase space density constraints place a lower bound on mass at around $\SIrange{1}{2}{\kilo\eV}$.
X-ray searches performed previously on different data sets are shown in the various parameter space plots \cite{Perez2017,HoriuchiHumphrey,BoyarskyMalyshev,NgHoriuchi,RiemerSorensen,NeronovMalyshev} for galactic and extragalactic observation regions. It is important to note however that these analyses are calculated using simulation-based DM halo profiles (see \cref{sec:RAR} for an in-depth comparison between the two approaches).

Bounds on the mixing angle $\theta$ can be placed due to sterile neutrino production mechanisms in the early universe, labeled in Figure 1 in \cite{Perez2017} as $\nu$MSM, BBN (lower bounds) and NRP (upper bound).
These mechanisms are heavily dependent of the values for lepton asymmetry in this early stage: in the absence of lepton asymmetry Non-Resonant Production (NRP) \cite{DodelsonWidrow94,AsakaLaineShaposhnikov,2017JCAP...01..025A} mechanisms are the only ones to take place, and by requiring the observed abundances to be produced a relation between $m_{s}$ and $\theta$ can be plotted in this case. The presence of lepton asymmetry adds another available production mechanism: Resonant Production (RP) \cite{LaineShaposhnikov,ShiFuller,AsakaLaineShaposhnikov}. Part of the DM abundance can be generated in this way, allowing the observed abundances to be met for smaller mixing angles. In the context of $\nu$MSM lepton asymmetry can be produced via decays of heavier sterile neutrinos, up to the value that outlines the line labeled as $\nu$MSM in Figure 1 in \cite{Perez2017} set by requiring this lepton asymmetry to be produced within the model. 
If we remain agnostic to the origins of this asymmetry, limits can be further lowered until these values come into conflict with nucleosynthesis predictions up to a value of lepton asymmetry of $L_6 \sim 700$. \cite{Boyarsky2009}.

These lower bounds can potentially be relaxed with the inclusion of self-interacting DM. Indeed, as is discussed in \cref{sec:interactingDM}, a novel extension to the $\nu$MSM paradigm via a vector boson interaction can provide a new channel for sterile neutrino production. In the presence of these new mechanism these lower bounds cease to be robust, so they are not included in \cref{fig:RAR_vs_P2017}. However it is extremely important to note that while these stringent bounds can potentially be avoided, a full thermal history of the sterile neutrinos under these self-interactions is not presented here and a more in-depth discussion of the subject is needed to provide a categorical conclusion.

As sterile neutrinos in $\nu$MSM tend to be produced non thermally with a non negligible free streaming length in the early universe, this affects the predictions for structure formation. From current cosmological data it is possible to constrain these models by performing analysis on observations related to the primordial power spectrum, such as from Lyman-$\alpha$ forest and substructure observations in the Local Group. Comprehensive reviews of the constraints for sterile neutrinos can be found in e.g. \cite{2017JCAP...01..025A,Boyarsky:2018tvu}
Conservative lower bounds on particle mass can be set using Tremaine-Gunn bounds on phase space density \cite{Tremaine:1979we,2017JCAP...01..025A}. This constraint lies in the range $\mathcal{O}(\SI{100}{\eV})$. In the case of halo counting and sub-structure observations it is difficult to establish robust predictions, but particles with masses $\gtrsim \SI{1}{\kilo\eV}$ are generally allowed (see for example \cite{Boyarsky:2018tvu} for a review of these problems). 

In summary, the parameter space for $\nu$MSM DM is heavily constrained by all the observations here mentioned: X-ray bounds, production mechanism limits and structure formation, as well as phase space considerations \cite{Perez2017,Boyarsky:2018tvu}. Here, we aim to study the effects on these constraints under the self interacting extension outlined in \cref{sec:interactingDM}: a new production mechanism can relax lower bounds on the mixing angle, while allowing for larger sterile neutrino masses (see \cite{deGouvea2019} for similar results). Besides, the inclusion of generalized RAR profiles (see \cref{sec:RAR}), can provide alternative observation targets for X-ray searches as discussed in \cref{sec:bounds}.


\subsection{Commonly used DM halo density profiles}
\label{sec:profiles}

DM density profiles for galaxy halos have been reviewed by many authors and is still a topic of discussion. As mentioned above,  since the expected photon flux from DM decays is proportional to the DM distribution inside the halo, the density profile plays a critical role in DM searches. 
One of the most commonly used parametrization is the NFW profile~\cite{Navarro:1995iw, Navarro:1996gj}
\begin{equation}
\rho_{\rm NFW}(r)=\frac{\rho_{\rm s}}{\left(r/r_{\rm s}\right)\,\left(1+\left(r/r_{\rm s}\right)\right)^2} ~,
\end{equation}
where $r_{s}$ is the scale radius and $\rho_{s}$ is the dark matter density at the scale radius. In this work, we consider the local DM density $\rho_{\rm local}=\rho_{\rm NFW}(r=r_{\odot}=\SI{8}{\kilo\parsec})= \SI{0.4}{\giga\eV/\centi\metre^3}$ and $r_{s}=\SI{21}{\kilo\parsec}$, which are compatible with the preferred parameters for the MW halo reviewed in \cite{ReadLocalDM}.

In order to study the impact of the density profile choice in the calculation of the $S$ factor, we consider other  alternative functional forms as the Einasto profile \cite{Einasto2006},
\begin{equation}
\rho_{\rm EIN}=\rho_{\rm s}\, {\rm exp} \left\lbrace -\frac{2}{\alpha} \left[ \left(\frac{r}{r_{\rm s}}  \right)^{\alpha}-1 \right] \right\rbrace ~,
\end{equation}
and the Burkert profile \cite{Burkert:1995},
\begin{equation}
\rho_{\rm BUR}(r)=\frac{\rho_{\rm s}}{\left(1+r/r_{\rm s}\right)^{\alpha}\,\left(1+r/r_{\rm s}\right)^{\beta}} ~,
\end{equation}
In this case, our choice is $r_{s}=\SI{21}{\kilo\parsec}$, $\alpha=0.17$ for the Einasto profile and $r_{s}=\SI{6}{\kilo\parsec}$, $\alpha=1$ and  $\beta=2$ for the Burkert profile. We also consider for both parametrization the same local DM density as described above.

While DM only simulations of the Milky Way favor profiles with density slopes similar to NFW at small radius, the scenario changes with the addition of baryons. Reference \cite{Calore}, which considered simulated galaxies with baryons that best fit the Milky Way data showed that the density slope is steeper for $\SIrange{1.5}{6}{\kilo\parsec}$, and shallower below $\SI{1.5}{\kilo\parsec}$ compared to NFW.
A conservative approach to this data is considering a density profile identical to NFW, but with constant density below the $\SI{1.5}{\kilo\parsec}$ range. We denote this profile as coreNFW. 

\subsection{RAR profile}
\label{sec:RAR}

The recently proposed RAR model \cite{2015MNRAS.451..622R} is based on a self-gravitating system of massive fermions at finite temperatures within general relativity, and its extension presented in \cite{Arguelles2016,Arguelles2018,amrr} accounts for the particle escape effect and galaxy finite-size via a particle energy cutoff in the phase-space distribution.
The solutions of the governing equations of this theory harbor a variety of different morphologies for the mass-distribution: for negative to low positive values of the central degeneracy parameter, $\theta_0 \lesssim 10$, the density spatial distribution shows a purely dilute isothermal-like profile where no fermion-degeneracy effects arise \cite{1990A&A...235....1G,2015MNRAS.451..622R}, and for $\theta_0\gtrsim10$, it appears a more general density profile with a compact \textit{quantum} core, supported by quantum degeneracy pressure, surrounded by an extended and diluted halo \cite{2015MNRAS.451..622R,Arguelles2016}.

The RAR profile with a dense and compact \textit{quantum} core has been shown to be of particular astrophysical interest since, besides explaining the MW outer halo rotation curves, it can also explain the dynamics of the innermost S-cluster stars around SgrA*, working as an alternative scenario to the central supermassive BH (see \cite{Arguelles2016} for details). From a structure formation point of view,
these \textit{core - halo} configurations can be shown to be reachable on cosmological timescales, being thermodynamically stable solutions that fulfill a maximization entropy principle at the end-state of its coarse-grained collisionless evolution \cite{daky}. The main underlying physical mechanism behind such a maximization, is the violent relaxation process, able to relax collisionless self-gravitating systems (either made of stars \cite{1978ApJ...225...83S}, or elementary fermions \cite{1996ApJ...466L...1K}), within dynamical timescales $t_D$ much shorter than the traditional collisional timescales, as demonstrated in \cite{1998MNRAS.300..981C}, complementing the original work of Lynden-Bell \cite{1967MNRAS.136..101L}.

The RAR profiles have distinct features that differentiate them from the phenomenological ones described in \ref{sec:profiles}. First of all, being based on a Fermi-Dirac phase-space distribution, they have an explicit dependence on the fermion mass\footnote{For indirect DM detection purposes, the particle mass is a parameter which only appears in the decay rate factor for previous profiles, thus making the $S_{\rm DM}$ factor not completely independent of the decay rate particle model within the RAR model.}.
Even if the main effects of such new physics are more pronounced through the inner part of the DM halo, they also imply specific consequences for the outer regions. The relevant point here is that the equation of state (Eqs.~\ref{eq:motionEq_19} and \ref{eq:motionEq_20}) intrinsically covers different physical regimes of a fermion gas, from classic Boltzmannian to quantum degeneracy. 
It is well-known that a fermion gas can transit in a continuous way in these regimes depending solely on the specific values of the fermion chemical potential and temperature. Since self-gravity and equilibrium require that the thermodynamical quantities change as a function of the gravitational potential, a specific link between the center and the outer parts of the system is created. This linkage allows the different fermion regimes to show up at different spatial scales, producing density profiles (see \cite{Arguelles2016,Arguelles2018} and  \cref{fig:rho-SIDM} and \cref{fig:vrot-SIDM} within the SIDM case analyzed here) with the following characteristic features (see \cite{Arguelles2016} for further details):

\begin{enumerate}
\item 
an inner core with radius $r_c$ (typically at mpc scales or below, for $m\sim 10-100$ keV particles) of almost constant density governed by quantum degeneracy; 
%
\item
an intermediate region with a sharply decreasing density distribution followed by an extended plateau, where quantum corrections are still important; 
%
\item
a King-like density tail governed by thermal pressure, showing a behavior $\rho\propto r^{-n}$ with $n>2$ due to the cutoff energy constraint.
\end{enumerate}

It is worth mentioning that the plateau $+$ halo tail behaviour following the quantum core, can vary from isothermal-like (when $\epsilon_c \rightarrow\infty$) to polytropic-like (for finite cutoff energy constraints $\epsilon_c$). For instance, see the $\rho\sim r^{-4}$ behaviour at the RAR halo scale-length $r_h$, shown in Fig. 6 in \cite{Arguelles2018}, in the application of the RAR model to different galaxy types, including the MW\footnote{Overall power-law behaviours like this have been shown to be typical of halo tails arising within numerical simulations aimed to test violent relaxation physics \cite{2009MNRAS.397..775J}. While overall halo tails behaving as $\rho \sim r^{-3}$ (typical of cosmological simulations), are expected to arise due to incomplete violent-relaxation and evaporation effects (see a discussion in \cite{2015PhRvD..91f3531C}).}. More generally, as shown in \cite{Arguelles2018}, the halo tail predicted by the RAR profile can be very similar to one of phenomenological profiles such as the Burkert or the cored-Hernquist ones, but it differs from the broken power-law of the NFW profiles.

Since we are working under ``cold enough'' fermion masses in the $10$--$100$~keV range (in agreement with large scale structure), it seems natural to ask why both RAR $\&$ NFW profiles do not coincide on typical inner-halo scales\footnote{Another contrasting example to the classical CDM simulations, besides the fermionic RAR case, are the modern numerical CDM simulations accounting for the quantum wave behaviour of the DM components \cite{2014NatPh..10..496S} (i.e. solving Schr\"odinger-Poisson equations). The corresponding density profiles show a solitonic core - extended halo behaviour which differs as well with respect to the NFW broken power-law trend.}.
However, this is not a well-posed comparison, since both approaches are not on equal footing regarding the underlying physics.
The RAR profiles 
are built upon Fermi-Dirac-like phase-space distribution
, while NFW profiles 
are obtained from an \textit{a posteriori} fitting procedure of the DM mass-shell distribution, from outside-to-inside, of classical CDM simulations results with limited spatial-resolution, and under certain virialization prescriptions. For the NFW profiles,  there is no knowledge as yet of the underlying physics (e.g. the precise equation of state) supporting/stabilizing the very central regions of the halo
(see e.g. \cite{2011MNRAS.415..225G}).

In the RAR case, the fermion mass constraints ($m\sim 10-100$ keV) obtained only from rotation curve observables, naturally find sterile neutrino (primordial) candidates~\cite{Asaka:2005,2015PhRvD..92j3509P}, in agreement with other recent cosmological constraints such Ly-$\alpha$ forest~\cite{2009JCAP...05..012B,2017JCAP...06..047Y}, CMB observations and small scale structure (see~\cite{2017JCAP...01..025A} for a review), among others. However, the fact that the RAR core compactness is sensitive to the particle mass (e.g. the larger the mass the smaller the core size for given $M_c$), only for \SIrange{48}{345}{\kilo\eV}, the core-halo RAR solutions can fit the Milky Way rotation curve data with the central DM core being compact enough to work as an alternative to the BH scenario in SgrA* \cite{Arguelles2016}. Instead for lower particle masses (e.g. between $\sim 10$ and $48$~keV) the core-halo RAR solutions can fit as well the rotation curve but the central core does not provide the BH alternative \cite{Arguelles2016}. Further calculations about exquisite orbit fitting of S2 star and the G2 cloud within the RAR model for $\sim 50$ keV mass (from GRAVITY collaboration data), together with a discussion about the radiative counterparts around SgrA* are given in \cite{bakrr}. 

 Concluding this section, and for the particle mass range of interest (\SIrange{48}{345}{\kilo\eV}), we will adopt in next the logic of point \textbf{(A)} given in \cref{sec:intro}. This alternative is detailed in \cref{sec:interactingDM}, where it is shown how the AMRR (RAR+SIDM-right-handed neutrino) model with DM particle masses $\sim 50$~keV can be in agreement with all the MW rotation curve data together with bullet cluster constraints.

\section{Parameter space relaxation \& dark sector interactions}
\label{sec:interactingDM}

{
The $\nu$MSM model \cite{Boyarsky:2018tvu,2017JCAP...01..025A} identifies right-handed sterile neutrinos with keV masses as DM particle candidates.  From production mechanisms, Ly-$\alpha$ and X-ray bounds \cite{Perez2017}, tight constraints can be obtained on the mass of the lightest sterile neutrino under pure $\nu$MSM assumptions, leaving an allowed range such as in \cref{constrnmsm}.

The constraints may be relaxed if (minimal) generalizations of the model are considered, assuming self-interactions among the particles, induced, for instance, by the exchange of massive vector particles in a dark sector of the model. Such an extension was first proposed in \cite{amrr} (and extended here for the more realistic RAR model accounting for escape-of-particles effects \cite{Arguelles2016}) in an attempt to explain the observed rotation curves (from center to periphery), as well as alleviate discrepancies between observations at galactic (small cosmological) scales and predictions based on numerical simulations based on the $\Lambda$CDM model (``small-scale Cosmology crisis'')~\cite{2017ARA&A..55..343B}. 
Interactions with pseudoscalars (axion-like excitations) in either visible~\cite{pilaftsis} or dark~\cite{dk,chuda} sectors of the pertinent theory have also been considered. In particular, in \cite{pilaftsis}, Yukawa type interactions of right-handed neutrinos with axion pseudoscalars have been proposed as a novel mechanism for generating a Majorana mass for the right handed neutrinos beyond seesaw~\cite{1977PhLB...67..421M,1980PhRvL..44..912M,1980PhRvD..22.2227S,1981NuPhB.181..287L,1979GellMann,1979Yanagida}, and in this sense they can also be included in the model of \cite{amrr} in addition to the vector interactions, thus significantly affecting the pertinent constraints.

The presence of such extra ingredients, entail also the important hint that dark matter may consist not only of a dominant component, but of several species, playing a different r\^ole at various scales. This relaxes significantly constraints on mixing parameters between the DM candidates and Standard Model matter arising from the requirement of avoiding overclosure of the Universe.
It is the purpose of this section to discuss briefly several such scenarios and how they modify/relax the constraints pertaining to the $\nu$MSM model.

\subsection{Vector Interactions among sterile neutrinos: update on the RAR+SIDM}
\label{subsec:vector}

We commence our discussion by considering the self-interacting DM model of \cite{amrr}. Here, we will review the basic ingredients of the model, and subsequently extend it for a more general distribution function accounting for the escape of particles. The relevant Lagrangian is given by:
\begin{equation}
{\mathcal L}={\mathcal L}_{GR}+{\mathcal L}_{N_{R\,1}}+{\mathcal L}_V+{\mathcal L}_{I}\,
\label{eq:vector_Ltotal}
\end{equation}
where
\begin{align}
{\mathcal L}_{GR} &= -\frac{R}{16\pi G}, \nonumber \\
{\mathcal L}_{N_{R\,1}} &= i\,\overline{N}_{R\, 1}\gamma^{\mu}\, \nabla_\mu\,N_{R\,1}-\frac{1}{2}m\,\overline{N^c}_{R\, 1}N_{R\,1}, \nonumber \\
{\mathcal L}_V &= -\frac{1}{4}V_{\mu\nu}V^{\mu\nu}+\frac{1}{2}m_V^2V_{\mu}V^{\mu},\nonumber \\
\label{vector_int}
{\mathcal L}_{I} &= -g_V V_\mu J_V^\mu=-g_V V_\mu \overline{N}_{R\, 1}\gamma^{\mu}N_{R\,1}\, ,
\end{align}
with $R$ the Ricci scalar for the metric background, which, for the purposes of \cite{amrr} which was the study of galactic profiles, it was assumed static and spherically symmetric: $g_{\mu \nu}={\rm diag}(e^{\nu},-e^{\lambda},-r^2,-r^2\sin^2\varphi)$; with $e^{\nu}$ and $e^{\lambda}$ functions only of the radial coordinate $r$, and $\varphi$ denotes the polar angle. The quantity $\nabla_\mu=\partial_\mu\, -\, \frac{i}{8}\, \omega_\mu^ {ab}[\gamma_a, \gamma_b] $ is the gravitational covariant derivative acting on a spinor, with $\omega_{\mu }^{ab}$ the spin connection; $m$ is the Majorana mass of the sterile neutrino, whose microscopic origin was left unspecified in \cite{amrr}.
The right-handed sterile neutrinos are denoted by $N_{R\,1}$ and the superscript $c$ over a spinor field denotes the charge conjugate, satisfying the Majorana four-spinor condition, ${\mathcal N}^c={\mathcal N}$ (see \cite{amrr} for further model details and properties).

For simplicity we assume  minimal-coupling of the vector field with the sterile neutrino current $J^\mu_V$ in the
interaction term ${\mathcal L}_{I}$ (\ref{vector_int}).  This current is conserved if decays of sterile neutrinos are ignored, as done in \cite{amrr}.

From \cref{eq:vector_Ltotal}, we obtain, respectively, the Einstein, Proca, and Majorana equations:
\begin{equation}
\label{eq:Einstein}
G_{\mu\nu} + 8\pi GT_{\mu\nu} = 0,
\end{equation}
\begin{equation}
\label{eq:RMF_10}
\nabla_\mu V^{\mu\nu} + m_V^2V^\nu - g_VJ_V^\nu = 0,
\end{equation}
\begin{equation}
\label{eq:Majorana}
\overline{N_{R1}}i\gamma^\mu \overleftarrow{D_\mu} + \frac{1}{2}m\overline{N^c_{R1}} = 0,
\end{equation}
where $G_{\mu\nu}$ is the Einstein tensor and $T_{\mu\nu}$ is the total energy momentum tensor. In the presence of the vector boson mediator, this tensor has two components: $T_{NR1}^{\mu\nu}$ and $T_V^{\mu\nu}$, each of which, in the perfect fluid assumption, is described by (with $\rho$ and $P$ the energy-density and the pressure):
\begin{equation}
\label{eq:RMF_12}
T^{\mu\nu} = (\rho + P)u^\mu u^\nu - Pg^{\mu\nu},
\end{equation}

Following \cite{amrr}, we work here under the relativistic mean field (RMF) approximation, which describes the system as corresponding to a static uniform matter distribution in its ground state. So, the vector boson field, as well as the source currents, are replaced by their mean values in this state, which, on account of space translational invariance, are independent of the spatial coordinates $\overrightarrow{x}$. This and the requirement of rotational invariance imply that no spatial current exists, and only the temporal component of the current is non zero: $\left\langle J_V^0 \right\rangle = \left\langle\overline{N_{R1}}\gamma^0N_{R1}\right\rangle = \left\langle N^\dagger_{R1}N_{R1}\right\rangle$. These expressions within brackets denote the finite number density of right-handed neutrino matter times the temporal component of the pertinent (average) velocity. From \cref{eq:RMF_10} we obtain directly the mean-field vector boson (notice that eqn. \cref{eq:Majorana} is identically 0 in the RMF approximation)
\begin{equation}
\label{eq:vectormeson}
V_0 = \frac{g_V}{m^2_V}J^V_0
\end{equation}
where $J^0_V = nu_0 = ne^{\frac{\nu}{2}}$ and $u_0$ is the time-component of the average future directed four velocity vector. From the last equation, recalling that we are working with a system comprising of a very large number $N$ of fermions in thermodynamic equilibrium conditions at finite temperature, we can assume, for the fermion number density:
\begin{equation}
n = e^{-\nu (r)/2}\left\langle\overline{N_{R1}}\gamma^0N_{R1}\right\rangle = {\frac{g}{(2\pi)^3}}\int_{0}^{\epsilon_c} f_c(p)d^3p
\end{equation}
In the last equality the integration is carried out over the momentum space up the cutoff energy $\epsilon_c$.

Here, we deviate from \cite{amrr} and introduce a more realistic model for $f_c(p)$ the Fermi-Dirac distribution function with a particle-energy cutoff in the numerator above which particles escape from the system\footnote{Such (quantum) phase-space funtion can be obtained as a quasi-stationary solution from a generalized Fokker-Planck equation for fermions including the physics of violent (collisionless) relaxation and evaporation, appropriate for non-linear structure formation \cite{2004PhyA..332...89C}.}, reading
\begin{align}
\label{eq:FDc-PS}
f_c (\epsilon\leq \epsilon_c) =
\frac{1-e^{(\epsilon -\epsilon_c)/kT}}{e^{(\epsilon - \mu)/kT} + 1}, \,\,\,\,\,\,\, f_c(\epsilon> \epsilon_c)=0.
\end{align}
Here, $\epsilon = \sqrt{p^2 + m^2} - m$ is the particle kinetic energy, $\mu$ is the chemical potential with the particle rest-energy subtracted off, $T$ is the temperature, $k$ is the Boltzmann constant, and m is the fermion mass, while $g = 2s + 1$ is the spin multiplicity of the quantum states taking to 1 for the singlet right handed majorana component $N_{R1}$ (viewed as a spin $+1/2$ fermion with one helicity state).

This ansatz for the (coarse-grained) fermion distribution function, generalizes the original self-interacting DM analysis done in \cite{amrr} with the standard Fermi Dirac one. The importance of such a generalization lies in the fact that $f_c(p)$ naturally provides DM halos bounded in radius (allowing for a redistribution of fermions with respect to the standard Fermi-Dirac case) which was shown (in the non-interacting case) to provide excellent fits with the galactic halos while  
being able to provide an alternative to the central BHs \cite{Arguelles2016,Arguelles2018,2019arXiv190509776A}. The aim of the phenomenological analysis described below, is, to repeat such a study in the case where DM vector boson interactions are included, and obtain the allowed range of values of the interaction coupling that fit the observational data.

In the following,  we include the thermodynamic equilibrium conditions for the system of semi-degenerate self-interacting fermions in general relativity, i.e. the Tolman and Klein conditions in the presence of an external (vector boson $V_0$) field, which read (see \cite{amrr} for further details)
\begin{equation}
\label{eq:tolman}
e^{\nu (r)/2} T = \rm{constant},
\end{equation}
\begin{equation}
\label{eq:klein}
e^{\nu (r)/2}(\mu + m + C_Vn) = \rm{constant},
\end{equation}
where we have defined $C_V=g_V^2/m_V^2$, and used \cref{eq:vectormeson} to derive \cref{eq:klein}. It is clear that for $C_V = 0$, the standard Klein condition is recovered.

Finally, the energy conservation of the particles along a geodesic, in presence of our external $V_0$ field, yields
\begin{equation}
\label{eq:cutoff}
e^{\nu (r)/2}(\varepsilon + m + C_Vn) = \rm{constant}.
\end{equation}

Thus, the full system equations for our self-interacting DM problem consist in the Einstein equations \cref{eq:Einstein} together with the above conditions \cref{eq:tolman}, \cref{eq:klein} and \cref{eq:cutoff}, given here in the following dimensionless manner
\begin{align}
	\frac{d\hat M_{\rm DM}}{d\hat r}&=4\pi\hat r^2\hat\rho, \label{eq:eqs1}\\
	\frac{d\theta}{d\hat r}&=-\frac{1}{2\beta}
    \frac{d\nu}{d\hat r}\frac{1+\frac{C_Vm^2}{4\pi^3}\left[\hat n - \beta\frac{d\hat n}{d\beta} - \theta \frac{d\hat n}{d\theta} - W \frac{d\hat n}{dW} \right]}{1+\frac{C_V m^2}{4\pi^3}\frac{1}{\beta}\left[ \frac{d\hat n}{d\theta} + \frac{d\hat n}{dW}\right]},\label{eq:eqs2}\\
    \frac{d\nu}{d\hat r}&=\frac{2}{\hat r^2} \left[\hat M_{\rm DM}+4\pi\hat P\hat r^3\right]\left[1-2\hat M_{\rm DM}/\hat r\right]^{-1}, \label{eq:eqs3}\\
    \beta(\hat r)&=\beta_0 {\rm e}^{\left[\nu_0-\nu(\hat r)\right]/2}, \label{eq:eqs4}\\
    W(\hat r)&=W_0+\theta(\hat r)-\theta_0\, .\label{eq:Cutoff}
\end{align}
In the limit $W\to\infty$ (i.e. $\epsilon_c\to\infty$) and $C_V=0$ this system reduces to the equations considered in the original RAR model \citep{2015MNRAS.451..622R}, while only taking $C_V=0$ the system leads to the more realistic version of the RAR model presented in \cite{Arguelles2016}. We have introduced the same dimensionless quantities as in the original RAR model formulation: $\hat r=r/\chi$, $\hat{n} = Gmn\chi^2$, $\hat M_{\rm DM}=G M_{\rm DM}/\chi$, $\hat\rho=G \chi^2\rho$, $\hat P=G \chi^2 P$, where $\chi=2\pi^{3/2}(\hbar/m)(m_p/m)$ and $m_p=\sqrt{1/G}$ the Planck mass. We have also introduced the temperature, the degeneracy and the cutoff parameter: $\beta = kT/m$, $\theta = \mu/(kT)$, and $W=\epsilon_c/(kT)$. We note that the constants of the Tolman and Klein conditions are evaluated at the center $r=0$, indicated with a subscript `0'.

In the presence of self-interactions, the total energy-density $\rho$ and pressure $P$ in \cref{eq:eqs1,eq:eqs2,eq:eqs3,eq:eqs4,eq:Cutoff} are given by the sum of the contributions of the energy-density and pressure of the fermions (in the RMF approximation) and of the vector boson mediator:
\begin{equation}
\rho = \rho_C + \rho_V,\quad P = P_C + P_V
\end{equation}
where the first component of the first equation is calculated as $\rho_C =  \left \langle T^0_0 \right \rangle _{N_{R1}}$ while $P_C = \left \langle T^1_1 \right \rangle _{N_{R1}}$. So, the total energy-density and pressure in presence of self-interaction are:
\begin{equation}
\label{eq:motionEq_19}
\rho = m\frac{1}{(2\pi)^3}\int_{0}^{\varepsilon_c}f(p)(1 + \epsilon/m)\,d^3p + \tfrac{1}{2}C_Vn^2
\end{equation}
\begin{equation}
\label{eq:motionEq_20}
P = \frac{2}{3}\frac{1}{(2\pi)^3}\int_{0}^{\varepsilon_c}f(p)(1 + \epsilon/m)^{-1}(1 +\epsilon/(2m))\epsilon\,d^3p + \tfrac{1}{2}C_Vn^2
\end{equation}
where $\rho_V=P_V=1/2 e^{-\nu}m_V^2V_0^2=1/2 C_Vn^2$ is the contribution from the vector boson field.

\subsection{Galactic phenomenology within the RAR+SIDM model}

The system of equations \cref{eq:eqs1,eq:eqs2,eq:eqs3,eq:eqs4,eq:Cutoff} form a system of coupled integro-differential equations, which must be integrated numerically for the following (regular) initial conditions at the center $r=0$: $M_{\rm DM}(0) = 0,\; \theta(0) = \theta_0,\; \nu(0) = 0,\; \beta(0) = \beta_0,\; W(0) = W_0$, for different DM particle mass $m$, and for appropriate coupling constants $C_V$, such that the $M_{\rm DM}(r)$ profile is in agreement with the observational constrains of DM halo rotation curve of the galaxy. For the latter we use the (general relativistic) formula for the circular velocity corresponding with a test-particle moving in the space-time metric here considered\footnote{This formula can be well approximated by $GM(r)/(r-2GM(r))$ as done in \cite{2015MNRAS.451..622R}, implying an error of less than $1\%$ for the quantum core compactness there considered.} (with $\nu(r)$ the $g_{00}$ metric function):
\begin{equation}
v (r) = \sqrt {\frac{d\nu(r)}{2\, d\ln{(r)}}}.
\end{equation}

\def\ROOTPATH{figure/rhoSidm}\begin{figure}
	\centering
	\includegraphics[width=0.9\hsize,clip]{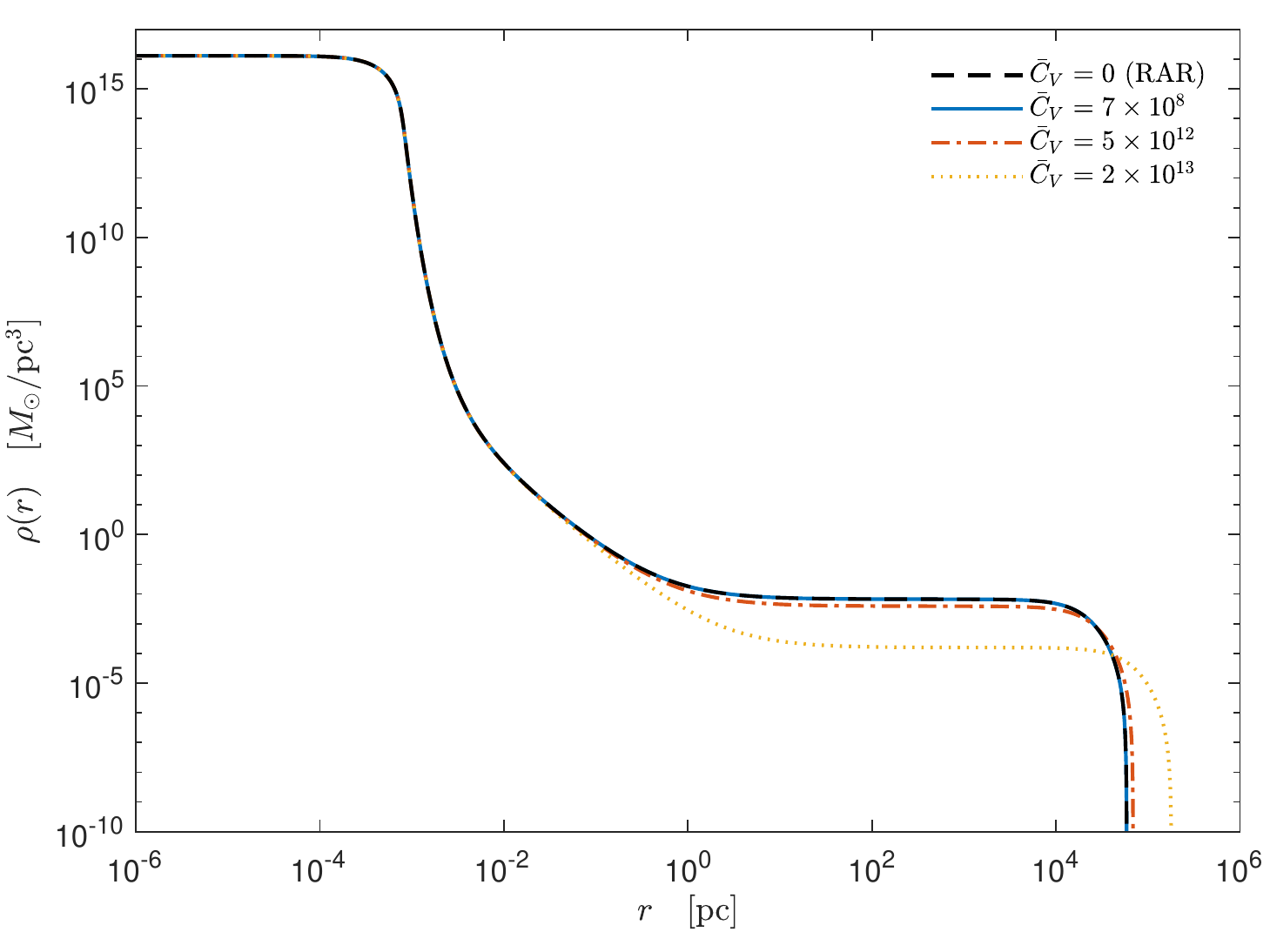}
	\caption{(Color online) Theoretical density profiles of the RAR$+$SIDM model including escape of particles and self-interactions. Shown are the solutions for four different coupling constant values, including the case without self-interaction ($\bar C_V = 0$, black dashed). Here the DM halo becomes more extended and diluted for higher $\bar C_V$ due to the additional pressure term from self-interactions.}
	\label{fig:rho-SIDM}
\end{figure}

\def\ROOTPATH{figure/vrotSidm}\begin{figure}
	\centering
	\includegraphics[width=0.9\hsize,clip]{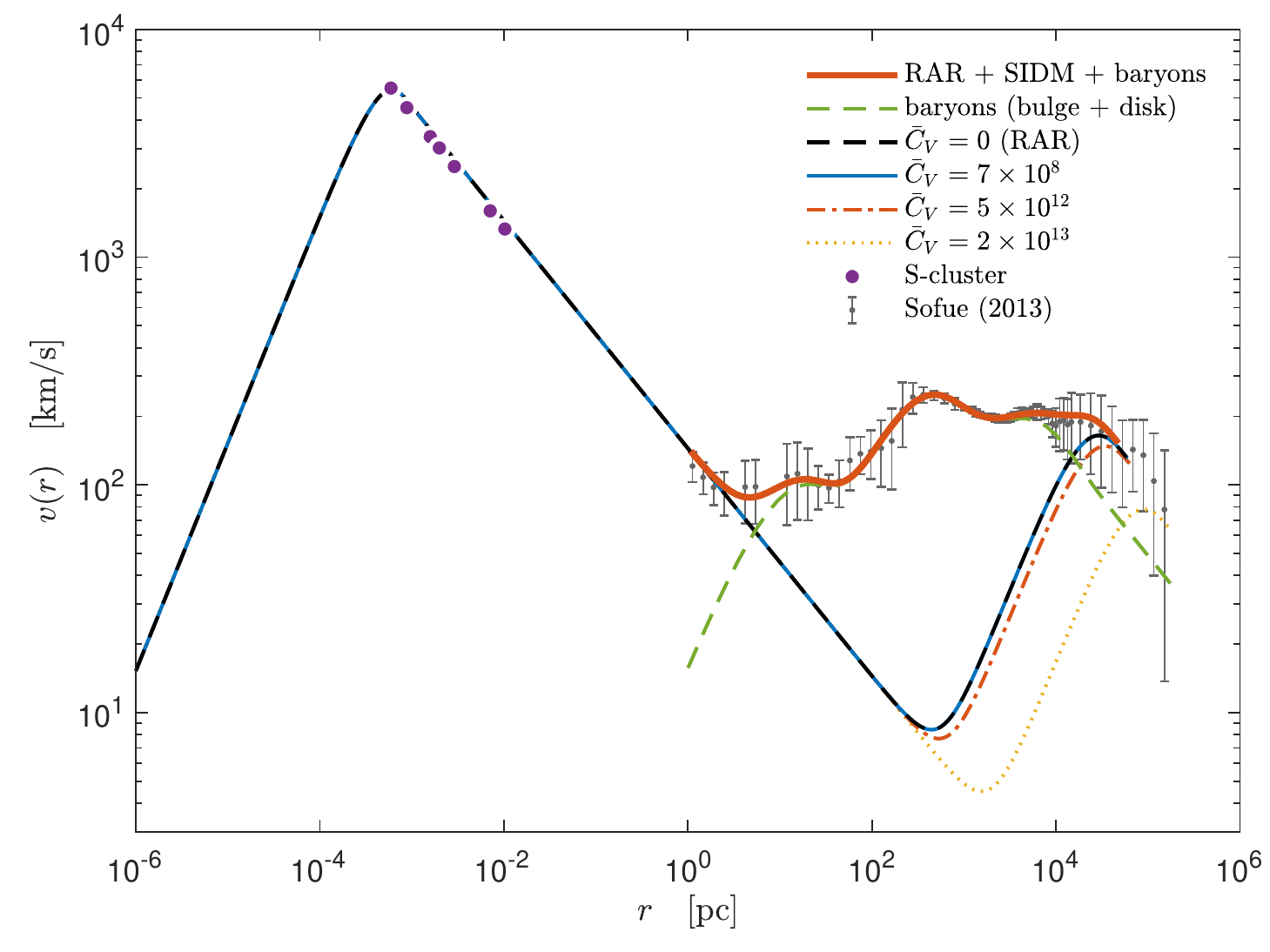}
	\caption{(Color online) Theoretical circular velocities of the RAR$+$SIDM model including escape of particles and self-interactions. Shown are the DM profiles for four different coupling constant values, including the case without self-interaction ($\bar C_V = 0$, black dashed). They are compared with data for the observed total rotation curve from \cite{2013PASJ...65..118S} (error bars). The theoretical total rotation curve (solid red), shows the excellent fit to the data for the RAR$+$SIDM composed by the DM with $\bar C_V= 7\times10^8$ (solid blue) and baryonic (green dashed) components (similarly to \cite{Arguelles2016}). Here the DM halo becomes more extended and less dominant for higher $\bar C_V$ due to additional pressure from self-interactions.}
	\label{fig:vrot-SIDM}
\end{figure}

We now give the numerical results for the DM density and rotation curve profiles, with the corresponding RAR+SIDM/AMRR parameter constraints) for the case of the Milky Way. In complete analogy with the phenomenological analysis done in \cite{Arguelles2016}, we adopt the following three boundary conditions to be fulfilled for DM halo mass $M_{\rm DM}(r)$, as inferred from the observations at two different radial locations in the Galaxy: a DM halo mass $M_{\rm DM}(r=\SI{40}{\kilo\parsec}) = \SI{2E11}{\Msun}$, consistent with the dynamics of the outer DM halo \cite{Gibbons14}, and $M_{\rm DM}(r=\SI{12}{\kilo\parsec}) = \SI{5E10}{\Msun}$, as constrained in \cite{2013PASJ...65..118S}. Simultaneously, we require a quantum core of mass $M_{\rm DM}(r=r_c)\equiv M_c = \SI{4.2E6}{\Msun}$ enclosed \textit{within} a radius $r_c=r_{p(S2)}=\SI{6E-4}{\parsec}$, the S2 star pericenter \citep{2009ApJ...707L.114G}. This implies three boundary conditions for the three free RAR model central parameters, once the particle mass and the interaction constant are specified.

In \cref{fig:rho-SIDM} and \cref{fig:vrot-SIDM} we show the density distribution $\rho(r)$ and the circular velocity $v(r)$ respectively. This is done for the relevant case of an ino mass of $mc^2 = \SI{48}{\kilo\eV}$, for different values of $\bar C_V \equiv C_V G_{\rm F}^{-1}$ with the fixed set of free-RAR parameters $\theta_0=37.1$, $\beta_0=10^{-5}$ and $W_0=65.2$, the latter fulfilling with the observable DM halo boundary conditions written above as demonstrated in \cite{Arguelles2016} for the non-interacting case $\bar C_V=0$ (in black-dashed here). Interestingly, we show here that there is a wide window of interacting constants, up to $\bar C_V\sim 10^{12}$ where no appreciable effects in the rotation curve appears (i.e. the contribution of the vector boson pressure to the total one is below $1\%$). We show explicitly in the plots the astrophysical value of $\bar C_V\sim 10^8$ in agreement with bullet cluster constraints as explained below (see \cref{eq:vector_5}). We further show that already for $\bar C_V\sim 10^{13}$ the additive vector boson contribution to the pressure starts to push forward the halo in an appreciable manner to the point to start spoiling the rotation curve fit (see continuous yellow line in \cref{fig:vrot-SIDM}). This completes (i.e. suffices within the scope of this paper) our phenomenological analysis on demonstrating the effects of the AMRR model on the DM halo of the Milky Way.}




Galactic phenomenology accounting for self-interactions (see ~\cite{amrr} and the more realistic RAR+SIDM extension developed here in the above subsection), implies modified Ruffini-Argüelles-Rueda (RAR) profiles (see~\cite{2015MNRAS.451..622R,Arguelles2016,siutsou,ars} for standard ones), with \textit{compact core-diluted halo} profiles, including for sufficiently dense cores in order to provide an alternative to the (SgrA*) massive BH in agreement with overall rotation curve observations {(for $\bar C_V\lesssim 10^{12}$). Such an agreement, shown in \cref{fig:vrot-SIDM} as an example}, can be acquired for a family of RAR solutions with a corresponding minimum mass of the right handed neutrino about $\SI{48}{\kilo\eV}$ (with its maximum allowed mass of about $\SI{345}{\kilo\eV}$, to avoid gravitational collapse) analogously as done in \cite{Arguelles2016}. 
Importantly, all these results have been obtained assuming a negligible mixing of the sterile neutrino with the SM sector. 

In addition, if one insists that the vector-field-induced self-interactions in the sterile neutrino sector provide solutions to the small-scale cosmology crisis~\cite{2017IJMPD..2630007M}, then a strong cross section relative to that of the conventional weak interactions in the SM sector, is required~\cite{amrr}. Indeed, to resolve such tensions between predictions of $\Lambda$CDM-based numerical simulations and observations, the self-interacting DM (SIDM) cross section has to be in the range~\cite{harvey}
\begin{equation}
0.1 \, \le \frac{\sigma_{\rm SIDM}/m}{{\rm cm}^2\, g^{-1}} \, \le 0.47~,
\label{eq:vector_newconstr}
\end{equation}
according to recent measurements employing novel observables of colliding galaxies (including the bullet cluster constraints). The vector interactions (\ref{vector_int}) in our case, lead to a cross section
\begin{equation}
\sigma^{\rm tot}_{\rm core}\approx\frac{(g_V/m_V)^4}{4^3\pi}29m^2 \qquad (p^2/m^2\ll 1)\, .
\label{eq:vector_20}
\end{equation}
which, on account of \cref{eq:vector_newconstr}, imply for the strength of the vector interaction of the sterile neutrino sector relative to the Fermi coupling $G_{\rm F}$ of the SM weak interactions~\cite{amrr}
\begin{equation}
\label{eq:vector_5}
\overline{C}_V \equiv \left(\frac{g_V}{m_V}\right)^2 G_{\rm F}^{-1} \in (\num{2.6E8},\num{7E8}),
\end{equation}
for ino masses in the range $m = \SIrange{48}{345}{\kilo\eV}$, implying that the mass of the massive-vector boson would be constrained to values $m_V\lesssim \SI{3E4}{\kilo\eV}$, in order to satisfy $g_V\lesssim 1$ as requested by the self-consistency of the perturbation scheme we have applied to compute the cross section by \cref{eq:vector_20}.

Such vector boson interaction ansatz (in the parameter range shown in \cref{eq:vector_5} consistent with bullet cluster constrains), do not have significant contribution to the shapes of the RAR$+$SIDM profiles as explicited in \cref{fig:rho-SIDM}--\cref{fig:vrot-SIDM}. However, as discussed in the following section these kinds of models may significantly alter the production and freeze-out (in) scenarios that result in the observed dark matter abundance. If one considered the extension to the $\nu$MSM model via the inclusion of these interactions, the production bounds can be significantly altered. Indeed in \cref{fig:RAR_vs_P2017} we have performed an indirect detection analysis for the AMRR-$\nu$MSM self-interacting extension model (or RAR$+$SIDM) with $m\geq 48$~keV. This results, allow us to consider a plausible scenario in which these interactions may relax the lower bounds on the active-dark sector coupling $\theta$ of the sterile neutrinos.

\subsection{Production mechanisms and $\nu$MSM parameter-space relaxation}

The presence of the vector-sterile-neutrino interaction term ${\mathcal L}_I$ plays an important role in the relaxation of the constraints of \cite{Perez2017}, since it implies an additional production channel for the DM sterile neutrino $N_{R\, 1}$ through the decays of the massive vector field $V_\mu$ in the early universe. So, sufficient production of DM may be guaranteed even if one ignores any coupling of sterile neutrinos with the SM sector, by assuming negligible Yukawa couplings $F_{\alpha\, 1}$ (see \cref{yuk_s2}). By providing an additional channel of sterile neutrino production by means of a vector boson decay (see Appendix \ref{Appx:Vboson_Decay}) which guarantees the right DM abundance, lower bounds on on the SM mixing angle $\theta$ can be relaxed (see \cite{deGouvea2019} for similar results).

Indeed, as discussed in Appendix \ref{Appx:Vboson_Decay}, the rate of decay (width $\Gamma_1$) of the vector Boson into a pair of identical Majorana particles (whose mass  $ \simeq \mathcal{O}(50)~\si{\kilo\eV}$
is viewed as negligible when compared to that of the
boson $V_\mu$, $m_V \simeq \SI{1E4}{\kilo\eV}$, according to the phenomenological analysis of \cite{amrr} in order to reproduce the observed galactic structure) is given approximately at tree level by
\begin{equation}\label{vnuwidth}
\Gamma_1   \simeq \frac{g_V^2} {48\, \pi}\, m_V~.
\end{equation}
Quantum corrections may affect this result, but will not be the focus of our brief discussion in this work. In models with more than one generation of right handed neutrinos coupled to the vector field there are extra contributions to the total width, which amount to a multiplication of the result in \cref{vnuwidth} by the number of right-handed neutrino flavours $N_f$ (usually $N_f=3$, like in the case of $\nu$MSM~\cite{nuMSM}).

The freeze-out temperature of the reaction is estimated by equating $\Gamma_1$ in \cref{vnuwidth} with the Hubble parameter $H$ of the Universe, $\Gamma_1 = H$. Assuming standard cosmology, in which there is radiation dominance in the Early Universe, the Hubble parameter is expressed in terms the temperature $T$ as~\cite{kolb}
\begin{equation}\label{hubble}
H=1.66 \, T^2 \mathcal{N}^{1/2} M_{\rm Pl}^{-1}~,
\end{equation}
where $\mathcal{N}$ is the effective number of degrees of freedom of all elementary particles and $M_{\rm Pl}$ is the reduced Planck mass. For a minimal extension of the SM, with only right-handed neutrinos and the background $B_0$, we may estimate $\mathcal{N} = \mathcal{O}(100)$ at temperatures higher then the electroweak transition. Equating \cref{vnuwidth} with \cref{hubble} we obtain for the pertinent freeze-out temperature, $T_{D}$,
\begin{equation}\label{freeze}
T_{D} \simeq 6.3\cdot 10^{-2} \frac{|g_V|}{\mathcal{N}^{1/4}}\sqrt{m_V \, M_{\rm Pl}}
\end{equation}
As discussed above, the requirement of alleviating the small-scale cosmology crisis via these vector-sterile-neutrino interactions requires~\cite{amrr} $m_V \simeq \SI{1E4}{\kilo\eV}$, with $g_V = {\mathcal O} (1)$; we then obtain from \cref{freeze} that $T_D = {\mathcal O}(10^8)~\si{\giga\eV}$, which yields the ball park of temperatures in which the sterile neutrino DM abundance is created in our interacting DM model.

The calculation of the sterile-neutrino thermal abundance at the freeze-out can be done as usual by the solution of the pertinent system of Boltzmann equations, or better out of equilibrium thermal field theory techniques (e.g. Kadanoff-Baym equations).
In general, one may end up with overproduction of sterile neutrino dark matter that would lead to overclosure of the Universe, unless the would-be freeze-out temperature of the vector bosons
lies above the reheating (or even preheating) temperature of the Universe. The latter is not known but it might be constrained by some CMB observations, with a lower limit lying in the range $\sim \numrange{20}{900}\ {\rm TeV}$~\cite{preheat1,preheat2}. We observe that in our simplified model the freeze-out temperature in \cref{freeze} is much higher than such lower limits of reheating temperature, and hence overproduction of warm sterile neutrino DM, through the decays of the vector boson, might be achieved. Other ways of avoiding overproduction of DM is via the dilution of the relic right-handed neutrino density by release of entropy through. e.g. decays of the heavier right-handed neutrinos (in models with more than one generation of sterile neutrinos) after their freezeout~\cite{rhnsu2}.

 Regarding the primordial abundance of the vector boson itself, we will assume it is sufficiently large to guarantee the proper DM abundance. This assumption is, at this stage, to be checked in the future as it depends on the microscopical model that underlies our phenomenological assumption: under our model \eqref{eq:vector_Ltotal} the vector boson only couples to right handed neutrinos and is a dark sector mode. 

The addition to \cref{eq:vector_Ltotal} of a Yukawa Higgs-portal term as explicited above in \cref{yuk_s2}, changes the situation drastically. Indeed, as we already discussed, upon considering such a coupling, one obtains the stringent X-ray 
constraints of the mixing angle and mass of $N_{R\, 1}$ depicted in \cref{fig:RAR_vs_P2017}, given that \cref{yuk_s2} implies decays of the heavy neutrinos $N_I \rightarrow \nu H$, where $H$ denotes the Higgs excitation field, defined via: $\phi = \langle \phi \rangle + H $. In such a case $J^\mu_V$ is \emph{not} a conserved quantity. However, in the context of $\nu$MSM, the lightest of the heavy neutrinos decay time is longer than the age of the universe~\cite{nuMSM}, hence the latter can be considered as stable for all practical purposes, thus playing the role of a DM component.

The thermal history of the Universe in the combined model where both the interaction term (\ref{vector_int}) and the mixing (\ref{yuk_s2}) is more complicated and we shall not present it here. However,
the Dirac Yukawa coupling  of the mixing term given by \cref{yuk_s2} for a $\si{\kilo\eV}$ sterile neutrino, of interest here, is sufficiently weak (as required by the seesaw scenarios~\cite{1977PhLB...67..421M,1980PhRvL..44..912M,1980PhRvD..22.2227S,1981NuPhB.181..287L,1979GellMann,1979Yanagida} of generating a light active neutrino mass in the SM sector) so it cannot bring the sterile neutrinos into thermal equilibrium above electroweak-scale temperatures. So, in the presence of our vector interactions with a freeze-out of order $10^8$ GeV as given in \eqref{freeze}, the Dirac Yukawa coupling will not play a dominant role in the sterile neutrino abundance. However, there exists the possibility of self interactions significantly contributing to this abundance when considering non thermal production mechanisms, such as Dodelson-Widrow non resonant production due to mixing with active neutrinos in the presence of a self interaction potential. This results open the possibility of matching the correct abundances for a wide range of parameters \cite{deGouvea2019}.

In summary, such a decay of a heavy vector Boson can provide another production channel for the sterile neutrinos and account for the observed DM abundance without the Dirac Yukawa coupling falling under the underproduction limits discussed in \cref{sec:comparison}. While these interactions can, however, lead to overproduction of sterile neutrinos we have calculated the production temperature ($T_D = {\mathcal O}(10^8)~\si{\giga\eV}$). This temperature lies well above reheating and overclosure can be avoided through several means as outlined in this section. Other production mechanisms are also affected by the presence of self interactions, which can also lead to the correct abundance for DM \cite{deGouvea2019}.


\section{Signal analysis}
\label{sec:signal}

So far we have introduced all the ingredients in order to perform the analytical calculation of the DM decay flux. By comparing the expected flux defined in \eqref{eq:S_factor_definition} with the observed one (i.e. such that $F^{\rm obs}_{\rm max}\geq F$, see \cref{sec:bounds} below) it is possible to place upper limits on the sterile neutrino mixing angles as a function of particle mass. In this way, an astrophysical region must be selected, task which is presented in this section from NuSTAR observations \cite{Mori2015}. The more stringent limits to the $\theta$ parameter will come from regions with a low upper limit on observed flux (i.e. few observed photons) and a high $S$ factor, or a high expected theoretical flux (i.e. high expected photons), as evidenced in \cref{eq:S_factor_definition}. A suitable selection for the observation region must then fulfill both conditions.

\subsection{Galactic Center}
\label{sec:SA_GC}

The main conclusions of this work arise when considering the observations (photon flux) coming from the innermost parsecs of the Galaxy. This observation is centered around G369.95-0.04, which is a Pulsar Wind Nebula candidate located at about less than $\SI{9}{\arcsec}$ away from SgrA*, which we identify as the geometrical center of all DM density profiles here adopted. The observation spans a circular region of $\SI{40}{\arcsec}$ around the centroid of G369.95-0.04. The observational data with corresponding spectra has been obtained by the NuSTAR instrument, as presented in \cite{Mori2015}.


The spectral analysis of this region shows a rich variety of X-ray sources in the $\SIrange{2}{50}{\kilo\eV}$ band, according to \cite{Mori2015}. Such features include SgrA*, G359.95-0.04, SgrA East, stellar winds, element lines and the CHXE,\footnote{
	Central Hard X-ray Emission. According to the detailed spectral study of two nearby intermediate polars and the CHXE by \cite{Hailey:2016}, the CHXE emission is likely an unresolved population of massive magnetic cataclysmic variables (CVs).
} among others.

This is an observation area filled with X-ray sources, and is expected to have a high observed photon flux. 
According to the criterion that we have mentioned above for good observation regions (low photon flux and high $S_{\rm DM}$), the inner parsecs of the Milky Way would be (for commonly used DM profiles) a non optimal region for this analysis. However, for the profiles we have analyzed in this work this is indeed not the case.
Some DM profiles show a significant density increase at the inner parsecs of the galaxy, which leads to a boost in $S_{\rm DM}$ factor for those regions. RAR+SIDM profiles analysed here are the best example, as the inner density spike accounts for most of the $S_{\rm DM}$ factor contributions, several orders of magnitude above other profiles for the same central area (as seen in following sections).
Such high $S_{\rm DM}$ factors can be enough to overcome the high observed photon flux on these regions and provide tight bounds for these types of profiles.

\subsubsection*{Line Flux Upper Limit}

In order to successfully obtain limits on the sterile neutrino DM parameter space from observations, it is necessary to determine a maximum X-ray flux that could have possibly been originated from DM decay.

Null-detection hypothesis has been tested by \cite{Perez2017} for the GC region within $\SI{E2}{\parsec}$ from SgrA*.\footnote{
	Namely, using the 0-bounce photon analysis for the GC data as also considered in this paper for comparison purposes.
} As this condition is independent from DM halo modeling and relative instrumental errors are unchanged, we assume the hypothesis to hold for the central few \si{\parsec} region as well.
 Moreover, this hypothesis is justified given the fact that a best fit to the total observed Flux including only astrophysical sources has been obtained \cite{Mori2015}, using the same instrument and within expected error bounds. A detailed analysis based on the modeling of all the known sources within the central \si{\parsec}, leading (or not) to the explicit null-detection conclusion is out of the scope of this paper.

Currently, observed X-ray spectra for the diffuse emission of the GC show features which can all be accounted for by the emission of astrophysical sources. Thus, unidentified decays in this band can only fall within the statistical uncertainies for the current measured spectra. We will outline here a method for estimating a maximal X-ray flux given by a hypothetical dark decay channel as it was done in \cite{HowToFindWDM} for a featureless region of the sky, and discuss its applicability here in \cref{subsec:discussion_MLF}.

Sterile neutrino decay modes can be approximated as monochromatic within the energy definition of current instrument, so the expected shape of a DM decay line in an observed spectrum would be a Gaussian peak with its width determined by the energy resolution of the instrument itself. The actual amplitude of the peak should be the maximal value allowed by the fit to the astrophysical sources, without ''spoiling'' the goodness-of-fit analysis (under a certain criterion). Roughly speaking, this maximal height can be estimated using the observational errors: A decay peak with maximal height should not exceed the local errors of the measurement. 

\begin{figure}
\centering
\includegraphics[width=0.92\hsize,clip]{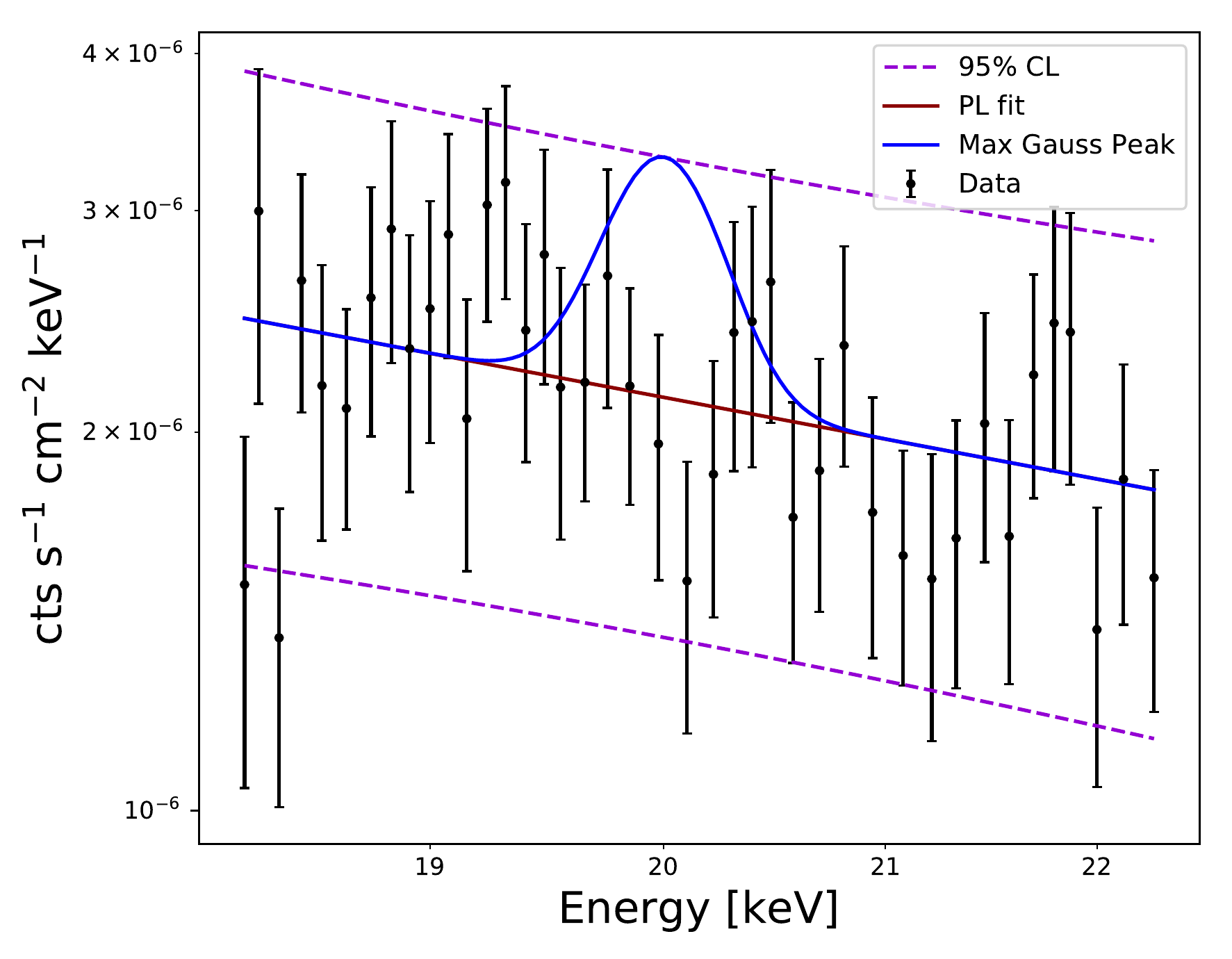}
\caption{An example of a line flux upper limit analysis for energy $\SI{20}{\kilo\eV}$. We have used the $\SI{40}{\arcsec}$ circular GC region diffuse emission spectra from \cite{Mori2015}. The black dots show an example data set, solid lines the power law (PL) and maximal Gaussian peak fits (red and blue respectively) with 95\% confidence bounds for the PL fit on dashed lines. }
\label{fig:Grafico_ej_maxflux}
\end{figure}

For sufficiently smooth spectra, the local errors can be calculated using a power law fit for a reduced spectrum around a given energy. The local error in the observation is then obtained from the 95\% confidence interval evaluated at the energy bin. Thus, an estimate for a maximal decay peak is obtained: its width determined by the energy uncertainties of the instrument, and its height by the 95\% confidence intervals of a local power law fit.\footnote{Local power law fits were defined in a $\pm$2 keV interval in order to reconstruct approximately the results in \cite{Perez2017} for 0-bounce photons: wider intervals overestimate flux around spectral lines.} An illustrative example of a peak centered around an arbitrary energy can be found in \cref{fig:Grafico_ej_maxflux}.

\begin{figure}
\centering
\includegraphics[width=0.92\hsize,clip]{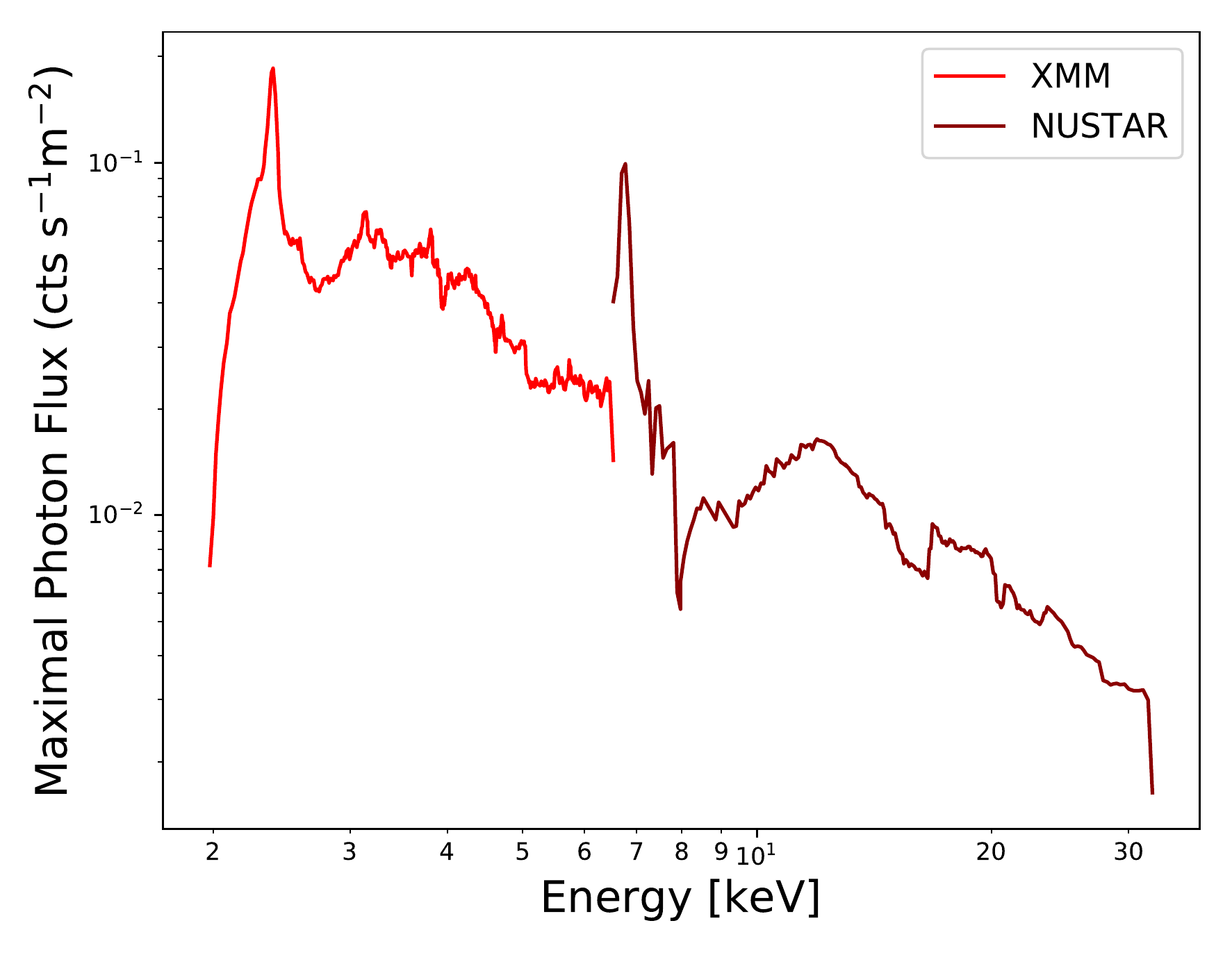}
\caption{Line flux upper limit for the XMM-NuSTAR observation, for $\SIrange{2}{30}{\kilo\eV}$ range X-ray particle decay. We used the  $\SI{40}{\arcsec}$ circular GC region diffuse emission spectra from \cite{Mori2015}.}
\label{fig:Grafico_maxflux}
\end{figure}

Such analysis has been performed for the diffuse X-ray emission spectra observed by the NuSTAR and XMM hard X-ray surveys \cite{Mori2015}. Making use of the spectra coming from a  $\SI{40}{\arcsec}$ region around the GC, we performed this analysis for a set of energies in the range $\SIrange{2}{30}{\kilo\eV}$ (due to sparse data in the $\SIrange{30}{50}{\kilo\eV}$ range) and plotted in \cref{fig:Grafico_maxflux} the line flux upper limit, calculated as the integrated flux of the maximal decay peak defined above.

The line flux upper limit suffers an enhancement around $\SI{6.5}{\kilo\eV}$, due to the power law approximation of the spectra failing around neutral Fe emission lines \cite{Koyama89,Koyama96}. \footnote{An ongoing discussion as to whether these lines correspond to DM decays or not is currently in progress, see for example \cite{Boyarsky:2018ktr}.} While this enhancement follows from the spectra not being locally well fitted by a power law due to astrophysical sources, it is also reflected in a degeneracy between these sources and the DM line in other constraining methods.

\subsubsection*{$S_{\rm DM}$ estimate}

From \cref{eq:exp_flux,eq:S_factor_definition}, the $S_{\rm DM}$ factor is obtained integrating over both the direction forming an angle $\theta$ with respect to the GC and along the line of sight, 
\begin{equation}
S_{\rm DM}=\int_{0}^{\infty}  \int_{\Omega_{\rm l.o.s.}}\rho(r(x,\theta))\,\d x \,\d\Omega ~,
\label{eq:S_factor}
\end{equation}
where  $r(x,\theta)=\sqrt{r_{\odot}^{2}+x^{2}-2r_{\odot}x \cos(\theta)}$. We developed a systematic process in order to calculate the $S_{\rm DM}$ factor  for each density profile considered in this work. Integration for central-cored profiles requires additional care since numerical processes yield inaccurate results for Dirac delta-like functions. The details about the calculation are discussed in \cref{appendix:SDM}. 

\begin{figure}
\centering
\includegraphics[width=0.92\hsize,clip]{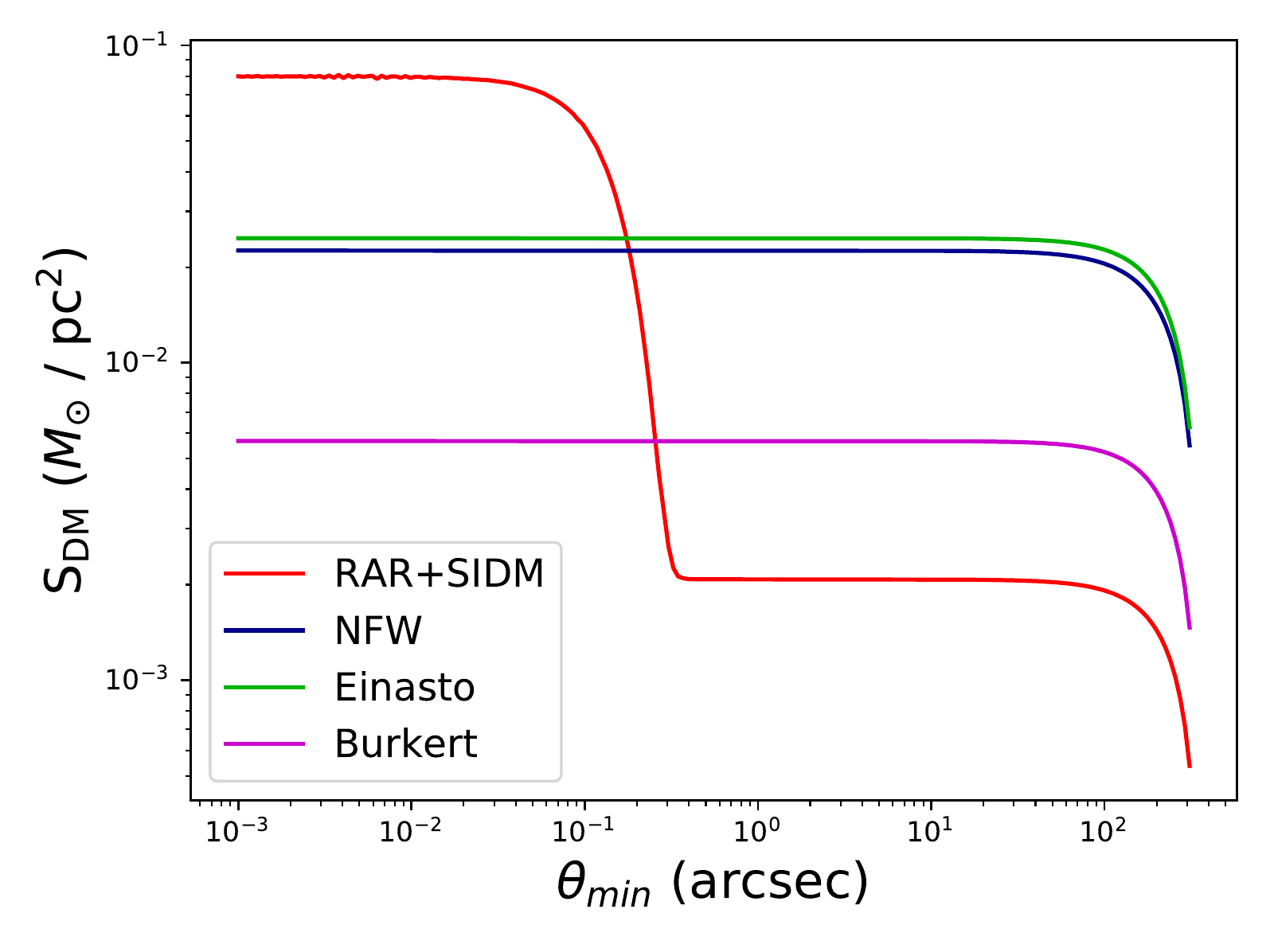}
\caption{$S_{\rm DM}$ factor for donut regions around the GC as a function of the region's minimum angle $\theta_{\rm min}$, for 4 different profiles. The maximum angle $\theta_{\rm max} = \SI{6}{\arcmin}$. NFW, EIN and BUR profiles use the parameter set specified in the text body, while the shown RAR+SIDM ($C_V=0$) profile uses a fermion mass $m = \SI{17}{\kilo\eV}$ and parameters fitting the MW halo (excluding S-stars) according to \cref{sec:RAR}.}
\label{fig:Grafico_comp_perfiles}
\end{figure}

Our results of the $S_{\rm DM}$ factor for the four different profiles are shown in \cref{tab:S_factor_gc}. We define the GC region as a $\SI{40}{\arcsec}$ circular area around the direction of the GC. Integration is performed on the full range of the $x$ coordinate along the line of sight.
\begin{table}
\centering
\begin{tabularx}{.45\textwidth}{Xc}
\toprule
\textbf{Profile Type} & $S_{\rm DM}$ $\left[\si{\Msun\parsec^{-2}}\right]$\\
\midrule
RAR+SIDM                & $\num{7.7904E-2}$   \\
NFW                   & $\num{3.6233E-4}$   \\
Einasto               & $\num{3.2055E-4}$   \\
Burkert               & $\num{6.9625E-5}$   \\
\bottomrule
\end{tabularx}
\caption{$S_{\rm DM}$ factor values for different profiles types with an integration region of $\SI{6}{\arcmin}$ circular area  around the GC. The parameters used for NFW, EIN, BUR profiles are specified in the \cref{sec:profiles}. In the case of RAR+SIDM profile we consider $mc^2 = \SI{17}{\kilo\eV}$ (and $C_V=0$) with the parameters adopted for the Milky Way halo as in Ref.~\cite{Arguelles2016}.}
\label{tab:S_factor_gc}
\end{table}
From these results is clear that the profile choice yields important differences in this factor. As evidenced in \cref{fig:rho-SIDM}, the RAR+SIDM profiles exhibit a boost of several orders of magnitude at a small radius (where $r < \SI{1}{\kilo\parsec}$). Thus, we expect a significant contribution in the $S_{\rm DM}$ factor due this small section. In order to quantify such contributions, we systematically calculate $S_{\rm DM}$ factors from `donut' shaped regions of the GC integrating from different $\theta$. We show the results for these example integration region in \cref{fig:Grafico_comp_perfiles} in order to clarify the contribution to the $S$-factor of the inner regions of the DM distribution. RAR+SIDM profiles were here calculated using a fermion mass $m = \SI{17}{\kilo\eV}$ and $C_V=0$ for the sake of example. Indeed, although this mass falls outside the range in which the DM quantum core of \textit{non-interacting} fermions offers an alternative to the BH scenario, it produces a density profile that shows all of the relevant features of the RAR profiles.

The $S_{\rm DM}$ factor (thus, the expected decay photon flux) undergoes boosting once the inner regions are included. Clearly from \cref{fig:Grafico_comp_perfiles}, if one neglected this region, the factor would become much smaller with respect to the case of other profiles, thus implying less stringent limits for RAR profiles within these observation target choices. 

\subsection{0-bounce photons}

A different observation region has also been considered, this time covering a much larger portion of the observed sky, with considerably smaller X-Ray flux. This observation follows the recent works by Perez and collaborators \cite{Perez2017} using the NuSTAR mission detectors, but aiming the analysis on the unfocused photons arriving at the detector without passing through the instrument’s focusing optics. When considering pointed observations of the GC, these photons account for the diffuse emission $\sim$~few $\SI{100}{\parsec}$ around SgrA*, however, vignetting effects due to physical blocking of the detectors by the instrument itself excludes up to the inner $\sim \SI{150}{\parsec}$, therefore reducing the astrophysical source contamination, but also removing the inner \si{\parsec} from the observation itself.

\subsubsection*{$S_{\rm DM}$ estimate}

For this analysis several observations are considered, roughly centered around the GC, following the procedures in \cite{Perez2017}. The total aperture from which these unfocused photons can reach the detector is about $\SI{3.5}{\degree}$ around the observation center, limited by the aperture stops attached to the focal-plane bench, and partially blocked by the NuSTAR instrument’s optics bench. These introduce both vignetting effects and physical blocking of photons arriving at the detector. Then, certain areas within the observation region are either completely blocked from detection, or the efficiency of the process is significantly diminished. Thus, to account for these effects, an efficiency factor is defined depending on the solid angle coordinates, and the S factor calculations are corrected for detector efficiency in the following form:
\begin{equation}
S_{\rm exp}=\int_{0}^{\infty}\int_{\Omega_{\rm FOV}}\epsilon(\Omega)\rho(r(x,\Omega))\d x \d\Omega ~,
\label{eq:S_factor_epsilon}
\end{equation}
with $\epsilon$ the detector efficiency factor ranging from 0 to 1.

The shape of the exposure maps for both X-ray detectors on board the NuSTAR mission are obtained in \cite{Perez2017,Wik:2014}. This sky-exposure correction factor takes into account vignetting effects and obscuration due to the instrument physically blocking photons from entering the detector from certain directions. The exposure map then excludes the inner parsecs of the GC for all observations considered here; a critical factor for dense core DM profiles as explained above.

As an example, we calculated these factors for a Field of View of 4 degrees around the GC, for the exposure map of detector FPMA for observation \texttt{obsID 40032001002}, for three different density profiles, obtaining results as in \cref{tab:S_factor_0b}. We include coreNFW profiles in the analysis following the arguments given in \cite{Perez2017}.

\begin{table}
\centering
\begin{tabularx}{.45\textwidth}{Xc}
\toprule
\textbf{Profile Type} & $S_{\rm DM}$ $\left[\si{\Msun\parsec^{-2}}\right]$\\
\midrule
RAR                & $0.3590$     \\
NFW                   & $2.3984$     \\
coreNFW               & $1.5041$     \\
\bottomrule
\end{tabularx}
\caption{$S_{\rm DM}$ factor values for different profiles types for NuSTAR mission data, using the methods described in \cite{Perez2017} for the 0-bounce photon region. The parameters used for NFW and coreNFW profiles are specified in the \cref{sec:profiles}. In the case of RAR+SIDM profile we consider, as an example, the case of $mc^2 = \SI{17}{\kilo\eV}$ ($C_V=0$) and best fit parameters for the MW halo excluding S-stars.}
\label{tab:S_factor_0b}
\end{table}

Due to the exposure map suppressing the contributions form the inner parsecs of the galaxy, RAR $S_{\rm DM}$ factors are significantly suppressed and remain under the ones obtained for NFW and coreNFW. These calculations for the S factor do not include, however, possible contributions from bad pixels or ghost rays (described in \cite{Wik:2014,Mori2015}, for example). These particular features however determine `bad data' regions and should be excluded from the observations and the S factor calculations. Both of these contributions can account for up to $\SI{70}{\percent}$ of the S factor, but are constant across profiles up to a $\SI{1.5}{\percent}$ standard deviation,\footnote{
	Tested for all profiles described in \cite{Perez2017}.
} thus remaining as an order of magnitude estimate and allowing us to provide comparisons between different dark matter profiles.

\subsubsection*{Line Flux Upper Limit}

The joint spectra from this analysis can be seen in \cite{Perez2017} for both detectors on board the NuSTAR mission: FPMA and FPMB, as well as an in depth analysis for these signals: a recount of the astrophysical sources considered in the spectral fitting and details on the spectral reduction methods.

\begin{figure}
\centering
\includegraphics[width=0.92\hsize,clip]{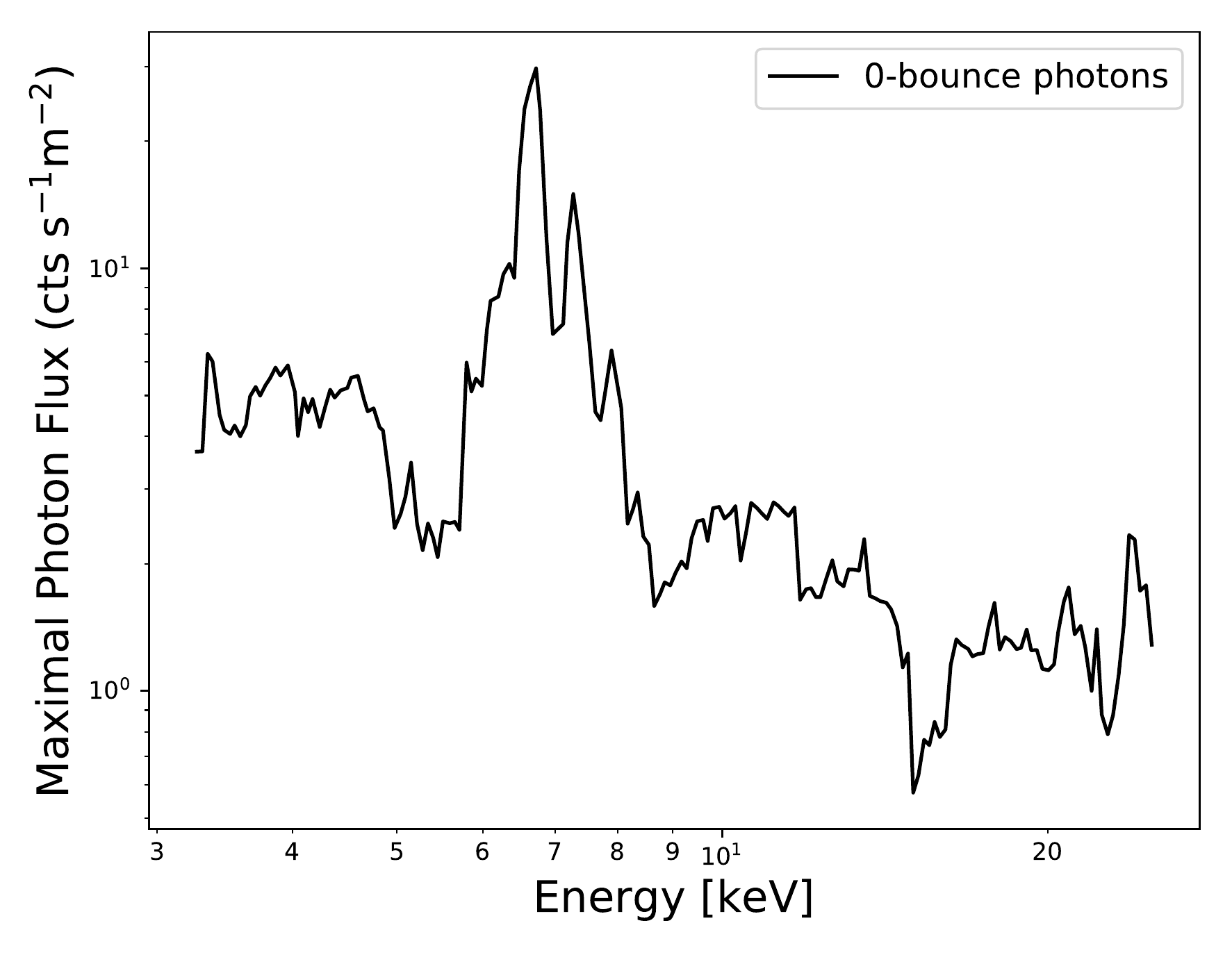}
\caption{Line flux upper limit for the 0-bounce photon observation, for $\SIrange{3}{25}{\kilo\eV}$ range X-ray particle decay. Used the emission spectra from \cite{Perez2017}.}
\label{fig:Grafico_maxflux_P2017}
\end{figure}

We have performed the line flux upper limit analysis for the added FPMA+FPMB spectra (normalized to the exposure time weighted averages of effective detector area and solid angle of sky coverage, see \cite{Perez2017} for details), and the results can be found in \cref{fig:Grafico_maxflux_P2017}. The expected flux is about a few $\si{\cts\ \second^{-1}\metre^{-2}}$, coming from a larger region of about $\SI{4}{\deg^2}$ total solid angle area. We chose to use a smaller energy range for this analysis than in \cite{Perez2017} to avoid areas where the detector background is the dominant spectral component. 

\subsection{On the estimation of the Line Flux Upper Limit}

\label{subsec:discussion_MLF}

We now discuss the applicability of the method for estimating the line flux upper limit from DM decays, as outlined in \cref{sec:SA_GC}. This procedure is most readily applicable for spectra with can be fit almost in their entirety by a power law  such as featureless diffuse emission (see \cite{HowToFindWDM}), and other common algorithms that are applied in the estimation of this quantity depend on the fitting to the astrophysical sources themselves \cite{Perez2017,Boyarsky2007}. These latter methods attempt to fit simultaneously all sources plus a DM peak component of unknown height, and establish its maximum value by constraining the deviation from optimal goodness-of-fit parameters. However, it is discussed in \cite{Perez2017} that the line flux upper limit obtained by this procedure can be roughly estimated by local observational errors, that are well reflected by the method used here. An argument can be made about the larger deviation between the two approaches on a section of the spectrum that could not be approximated well by a power law, however local error overestimation in the method used here is reflected in a degeneracy between astrophysical sources and the DM line in a fit-dependent algorithm.

It is important to state, however, that this method of flux constraining has to be taken as an \textit{approximation} to a full fit-dependent procedure as used on other works. The difference between the two can be quantified for the observation of 0-bounce photons, and the results of this approach on the parameter space limits can be seen in \cref{fig:Grafico_profile_compare_P2017}. While the results must be taken in light of this approximation, this does \textit{not} alter the conclusions on this paper in a significant way. 

Indeed, for the results on this region, we can outline quantitative differences between this method and other fit-dependent ones by comparing with the results on line flux upper limit obtained by \cite{Perez2017}. A $\SI{28.7}{\percent}$ mean difference excess between methods was observed within the full $\SIrange{3}{25}{\kilo\eV}$ spectral range. 

It is important to further stress that an overestimation of the observed line flux leads to a relaxation in observational limits (which follows directly from \cref{eq:theta_upper_bound}) and can only result in more conservative limits for the mixing angle $\theta$. Thus, if our limits to the parameter space using RAR profiles (as exposed in \cref{fig:RAR_vs_P2017}) would have been obtained using source-fit dependent methods in the analysis, it would lower the upper bound due to the method difference mentioned above by an average of $\sim \SI{30}{\percent}$ for this data set.

Thus, this method of flux constraining, while it does not depend on spectral fitting models, slightly overestimates the upper bounds when comparing with other fit-dependent methods by up to a factor of a few.

\section{Parameter space bounds}
\label{sec:bounds} 
\subsection{Galactic Center}

Having established an upper limit on the sterile neutrino decay flux, and having calculated the theoretical expected flux, it is straightforward to obtain a $(m_{s},\theta)$ parameter space limit. Claiming that the expected flux from DM decay, \cref{eq:exp_flux}, cannot exceed the upper limit from X-ray observations (i.e. we assume the null-detection hypothesis for this region):
\begin{equation}
F^{\rm max}_{\rm obs}\geq F=\frac{\Gamma}{4\pi m_{s}}S_{\rm DM} ~.
\label{eq:max_flux_condition}
\end{equation}
Recalling the expression for sterile neutrino decay rate given in \cref{eq:decay_rate}, a bound on the mixing angle $\theta$ can be obtained as a function of $m_{s}$ as:
\begin{equation}
\theta^2 \leq \num{1.9465E-4}
	\left [ \frac{F^{\rm max}_{\rm obs}}{\si{\ph\ \second^{-1}\centi\metre^{-2}}}\right ]
	\left [ \frac{\si{\kilo\eV}}{m_{s}}\right ]^{4}
	\left [ \frac{\si{\Msun\parsec^{-2}}}{S_{\rm DM}}\right ]
\label{eq:theta_upper_bound}
\end{equation}

This X-ray bound becomes more stringent as more accurate constraints on maximum observed flux are achieved. Thus, non observation of DM decay lines on higher resolution equipment or tighter analytical constraints on observed flux can only contribute to lower the bounds here obtained. The bound is also inversely proportional to $S_{\rm DM}$, so to obtain tighter constraints it is necessary to identify observational targets with boosted $S_{\rm DM}$ factors for a given DM profile. 

\begin{figure}
\centering
\includegraphics[width=1.0\hsize,clip]{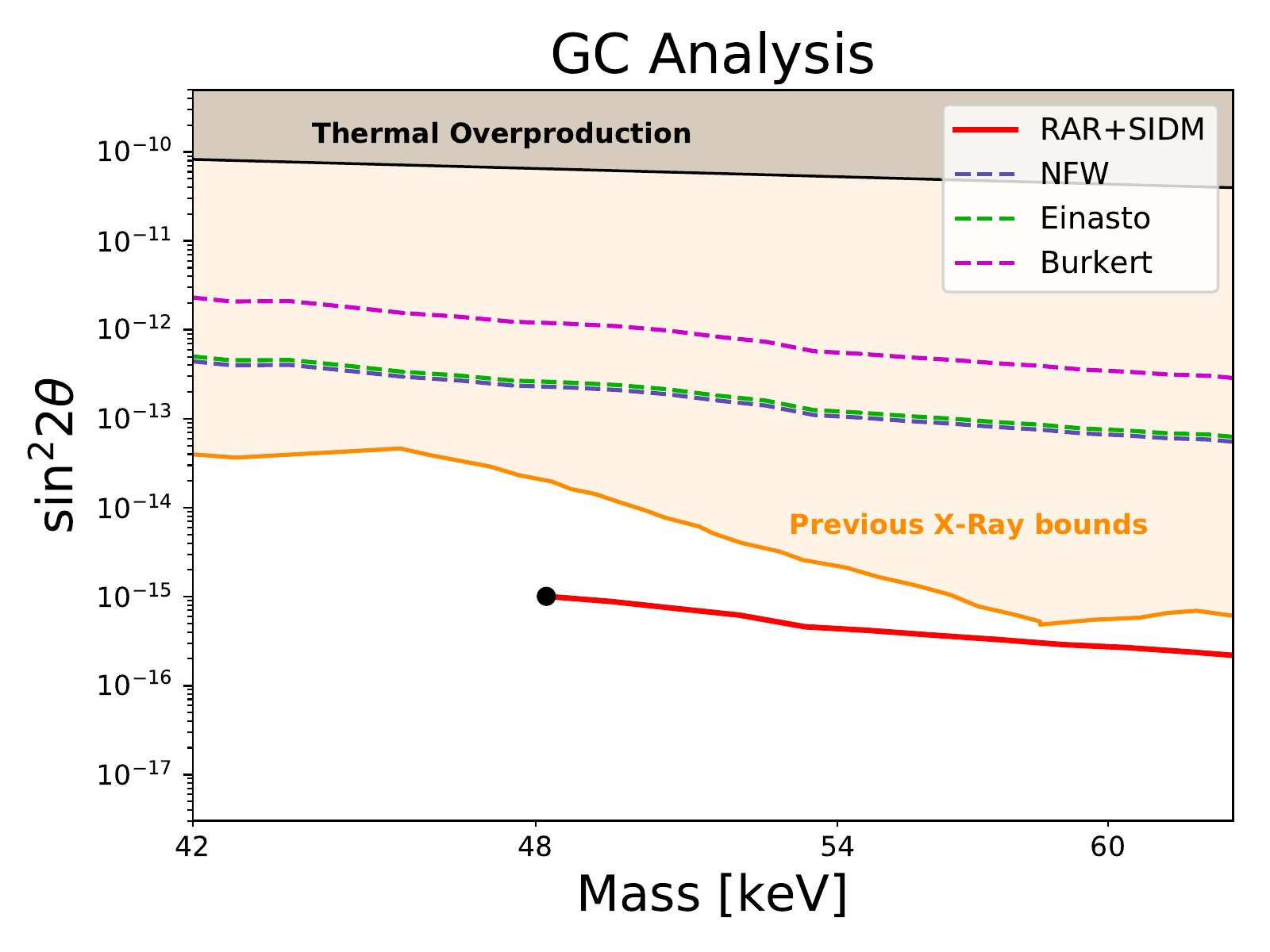}
\caption{ Sterile neutrino parameter space limits obtained for GC X-ray emission analysis. Four profile types compared: BUR, EIN, NFW and RAR+SIDM. The light Red shaded region above the continuous red line corresponds to RAR$+$SIDM limits from this work given by X-ray bounds (i.e. indirect detection analysis), while 
the black dot at \SI{48}{\kilo\eV} labels the smallest DM particle mass compatible with S-cluster stars' rotation curve data that can provide a BH alternative (see \ref{sec:RAR}). 
RAR+SIDM interaction strength is in line with \cref{eq:vector_5}, $C_V \sim 10^8$ in this plot.
The upper shaded region corresponds to production mechanism bounds: NRP under no lepton asymmetry. Lower bounds on $\theta$ are dismissed due to alternative production mechanisms with respect to standard $\nu$MSM.
Orange line marks the combined constraints of previous X-ray searches \cite{Perez2017,HoriuchiHumphrey}, see \cref{fig:RAR_vs_P2017} for individual constraints.}
\label{fig:Grafico_profile_compare}
\end{figure}

We have obtained these bounds for the mixing angle ($\theta$), and for the profiles mentioned above: NFW, Einasto, Burkert and RAR+SIDM. Results can be seen in \cref{fig:Grafico_profile_compare}. Analysis has been performed for the full mass range allowed by the spectra in the case of NFW, EIN and BUR profiles.

As expected, the $S_{\rm DM}$ factor enhancement for the novel RAR type of profiles results in 
the most stringent upper limits for this observation. As seen before, this enhancement results from the inclusion of the central regions of the GC in the observation, which increases $S_{\rm DM}$ factors and brings them over the ones arising from the other three profiles.

\subsection{0-bounce photon analysis: comparison with recent works}

We base our comparison mainly with the work by Perez et al. \cite{Perez2017}, who conducted searches using 0-bounce photons lowering the upper mixing angle bounds, and further narrowed down the allowed particle mass window than in previous works. The parameter-space bound analysis for the 0-bounce photons spectrum is very similar to the one previously mentioned. A few differences reside in the calculation for the $S_{\rm DM}$ factors.

Namely, the main differences reside in the exposure map corrections mentioned in previous sections, in addition to averaging over different observations. As the spectrum has been averaged over six observations, and co-added for FPMA and FPMB, each with different exposure maps, the expected flux must be obtained via a weighted average of $S_{\rm DM}$ factors for each one of the observations.

The $S_{\rm DM}$ factor has been calculated as:
\begin{equation}
S_{\rm avg}=\frac{\sum_{i} \Delta t \Delta \Omega S_{i}}{\sum_{i}\Delta t \Delta \Omega} ~.
\label{eq:S_factor_average}
\end{equation}
With $\Delta t$ the exposure time and $\Delta \Omega$ the effective detector area for each observation. The specific values of these parameters and further observation details can be found in \cite{Perez2017}.

\begin{figure}
\centering
\includegraphics[width=1.0\hsize,clip]{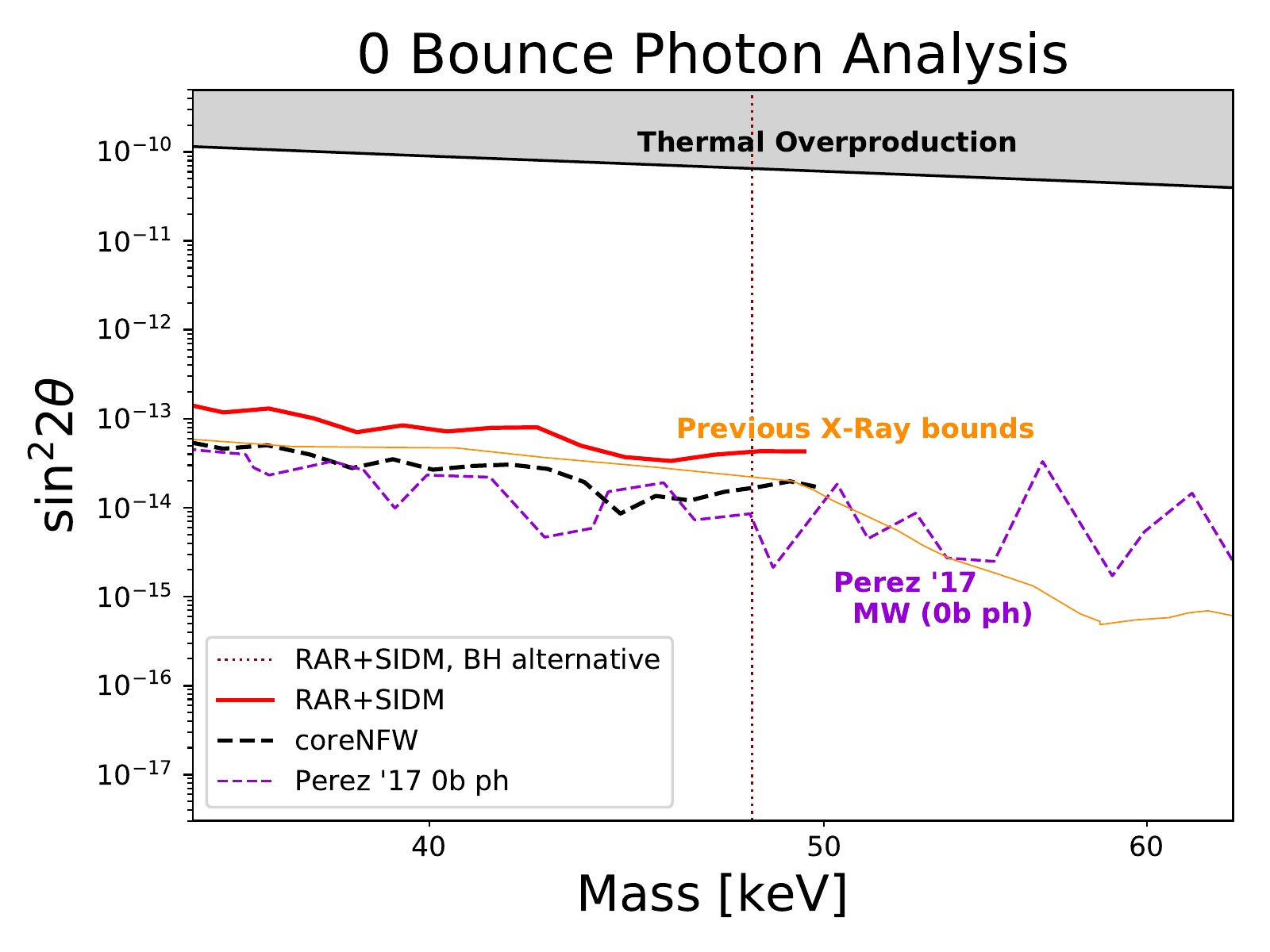}
\caption{ Parameter space limits obtained for 0-bounce photon ("0b ph") X-ray analysis. Two profile types compared: coreNFW and RAR+SIDM. Violet dotted line refers to analysis by \cite{Perez2017}, corresponding to a coreNFW profile. 
The vertical dotted line at \SI{48}{\kilo\eV} labels the smallest DM particle mass compatible with S-cluster stars' rotation curve data that can provide a BH alternative (see \ref{sec:RAR}). 
RAR+SIDM interaction strength is in line with \cref{eq:vector_5}, $C_V \sim 10^8$.
The upper shaded region corresponds to production mechanism bounds: NRP under no lepton asymmetry. Lower bounds on $\theta$ are dismissed due to alternative production mechanisms with respect to standard $\nu$MSM.
Orange line marks the combined constraints of previous X-ray searches \cite{Perez2017,HoriuchiHumphrey}, see \cref{fig:RAR_vs_P2017} for individual constraints.
}
\label{fig:Grafico_profile_compare_P2017}
\end{figure}

Once these averages have been taken, the procedure for obtaining a bound for $\theta$ are similar to the one taken for the GC. We performed the analysis for profiles coreNFW and RAR+SIDM (with parameters previously mentioned), obtaining the results in \cref{fig:Grafico_profile_compare_P2017}.

We remind the reader that the last data point in this plot corresponds to the region in which the source spectrum starts to be dominated by instrument noise, which we have chosen not to plot here. However, results for RAR profiles with masses $\leq 48$ keV (compatible with rotation curves, but not with data from S-cluster stars) allow us to draw conclusions about the generalities of the observation region. 
The limits for the RAR+SIDM profile are significantly relaxed for this region. This is to be expected, as inner compact regions for these profiles are excluded from the exposure map, therefore not contributing to the $S_{\rm DM}$ factor and thus relaxing the expected bounds.

\begin{figure}
\centering
\includegraphics[width=1.0\hsize,clip]{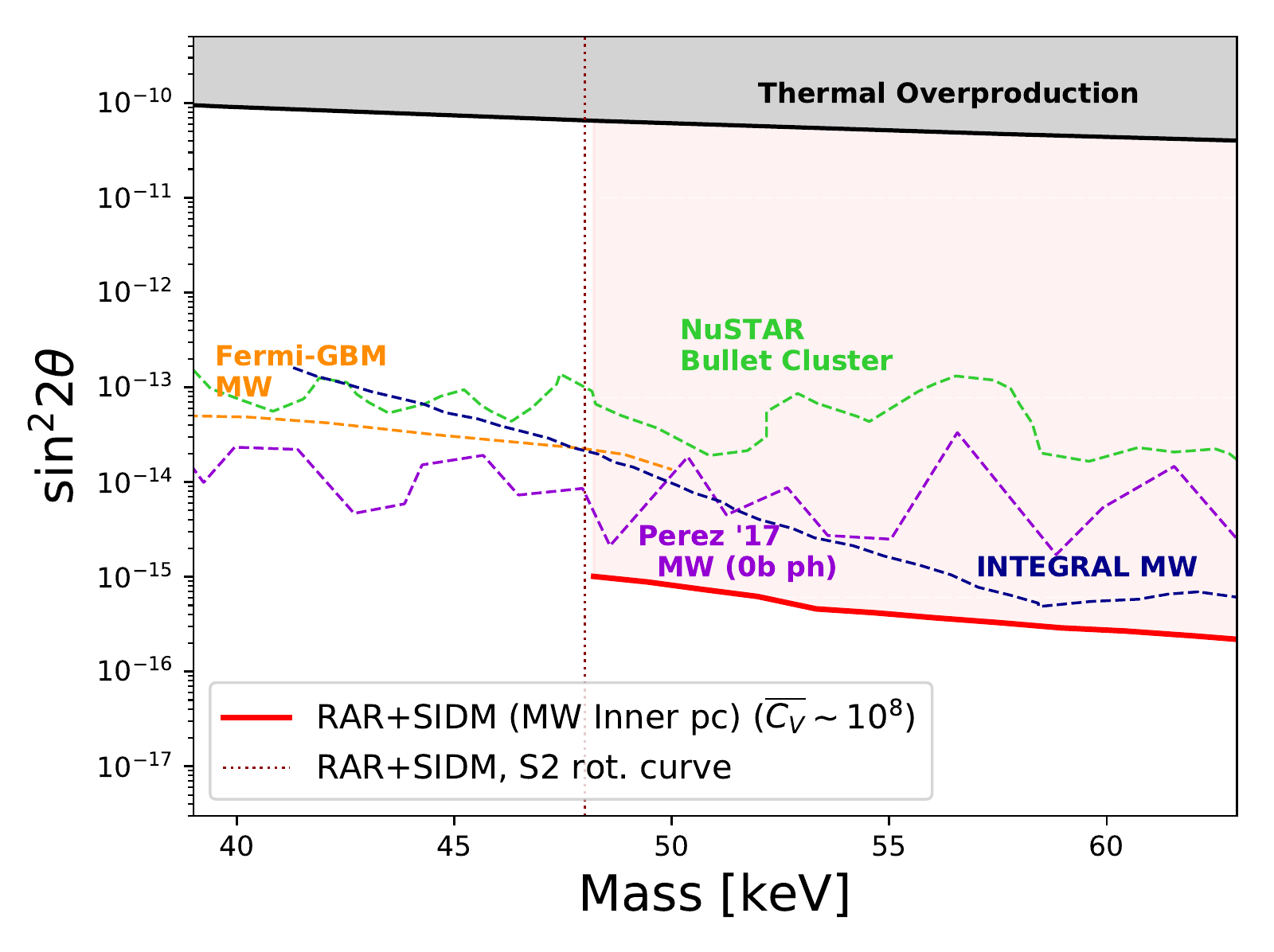}
\caption{Sterile neutrino parameter space limits obtained for GC observations using RAR$+$SIDM profiles (continuous red line), when assuming DM production due to self interactions (see Appendix B) and interaction strength in line with \cref{eq:vector_5}.
The light Red shaded region above the continuous red line corresponds to RAR$+$SIDM limits given by X-ray bounds (i.e. indirect detection analysis), while 
the vertical shaded region below \SI{48}{\kilo\eV} labels the smallest DM mass compatible with S-cluster stars' rotation curve data that can provide a BH alternative (see \ref{sec:RAR}). 
The upper shaded region corresponds to production mechanism bounds: NRP under no lepton asymmetry. Lower bounds on $\theta$ are dismissed due to alternative production mechanisms with respect to standard $\nu$MSM.
Other dotted lines refer to several X-ray bounds for different DM halo profiles including 0-bounce photon analysis \cite{Perez2017} in dashed violet (further discussed in \cref{sec:comparison}).}
\label{fig:RAR_vs_P2017}
\end{figure}

Summarising, we showed that for such a 0-bounce photon analysis, no stronger limits could be obtained from assuming RAR+SIDM profiles. However, for observations that include the inner parsecs of the GC, we found that RAR+SIDM profiles provide (slightly) stronger limits in the interaction angle than by using NFW or similar distributions. We then compare the limits obtained for GC observations with the limits obtained by \cite{Perez2017} in \cref{fig:RAR_vs_P2017}.

Two main Milky Way observations/DM profile pairs are shown: (i) expected fluxes corresponding to flat core NFW profile (motivated by observations, see~\cite{Perez2017}) are compared to NuSTAR observation for diffuse light $\sim \SI{100}{\parsec}$ around the GC, as reviewed in~\cite{Perez2017}, placing an upper limit on the sterile neutrino mixing angle (i.e. violet dashed line). In this work, observations of the very inner $\sim$~few \si{\parsec} of the Milky Way were used instead (see \cref{sec:SA_GC}), and compared against the RAR+SIDM model. An estimate for this upper bound is obtained (i.e. continuous red line), which turns out to be in line with previous bounds \cite{Perez2017}-but stronger- for these novel profiles, but only when central \si{\parsec} observations are included in the analysis.

These bounds are shown for masses $m_s \geq 48$ keV, the minimum mass value for RAR+SIDM profile compatible with S2 star rotation curve data. However the region of smaller masses is not excluded but simply unconstrainable with current DM halo profiles under discussion (see footnote 2). A robust upper particle mass bound under these model assumptions (i.e. within low enough coupling constant) is $m = 345$ keV as further commented in \cref{sec:RAR}. Further constraining the fermion mass range to lower values, could be achieved under a more complete thermal history of the particles, as we discuss in \cref{sec:comparison}, \cref{sec:interactingDM}.

Under the self-interaction paradigm explained in \cref{sec:interactingDM} the production mechanism bounds mentioned in \cref{sec:comparison} are indeed significantly relaxed as the decay of heavy vector bosons can produce sterile neutrinos in significant quantities. Thus, the lower bound on interaction angle $\theta$ in \cite{Perez2017}, which arises from the maximal allowed values for lepton asymmetry, is no longer robust. As we mention in \cref{sec:comparison} though, a further analysis into the thermal history of this model is required.
 
\section{Conclusions}
\label{sec:concls}

Considering a DM profile that self-consistently accounts for the particle physics model, we performed a new analysis of NuSTAR X-ray data to study how the $\nu$MSM parameter-space constraints are affected by self-interactions among sterile neutrinos. In particular, we have shown how standard production mechanisms in the early Universe can be affected through a decay of the massive vector field that acts as the mediator of the self-interactions of the DM candidates, which could broaden the allowed parameter space with respect to standard $\nu$MSM.

The novelty of the present paper lies in the fact that 
\begin{itemize}
\item We work within a particle-based DM halo model which derives self-consistently the particle mass dependent RAR+SIDM fermionic density profiles (see  \cref{sec:interactingDM}), unlike the ones previously utilized in the literature (e.g.~\cite{Perez2017}). Importantly, such RAR+SIDM profile is compatible with measurements of the Galaxy rotation curve as well as constraints on the DM self-interacting cross section from the Bullet cluster.a 
\item We use a new observation region within the Galaxy central \si{\parsec} taken from \cite{Mori2015}, where an indirect detection analysis of the X-ray NuSTAR data is performed. The constraints derived here on the self-interacting sterile neutrino parameter space, are stronger than those obtained with commonly used DM profiles, due to the dense DM quantum core which is characteristic of the RAR and RAR+SIDM profiles. 
\end{itemize}

Specifically, we have performed a null-detection analysis based on an expected signal from the lightest sterile neutrino decay in the X-ray energy range $E_{\gamma} \sim \SIrange{2}{50}{\kilo\eV}$, focused towards the inner parsecs of the Galaxy. Several indirect detection analyses on the traditional, non-self-interacting $\nu$MSM model, have been performed (see e.g. \cite{Perez2017}) assuming the traditional phenomenological DM halo profiles that arise from the fitting of simulation results with finite spatial resolution down to $\sim \SI{0.1}{\kilo\parsec}$ scales (or larger in the case of cosmological simulations). Thus, when aiming to diffuse emission regions inside the few pc radius, extrapolated versions of those profiles are used, due to the lack of knowledge of the DM content around SgrA* (see section \ref{sec:RAR} for further discussions).

It has been thus one of the main motivations of this work to provide consistent fermionic (``ino'') halo models (i.e. the RAR \cite{Arguelles2016}/ RAR+SIDM profiles) for such a detection analysis, where a precise knowledge of the DM distribution from the center to periphery is available from first principle physics (see \cref{sec:RAR} for details). We performed such tests upon the self-interacting AMRR extension of the $\nu$MSM model~\cite{amrr}, which allows in principle for a significant relaxation of the constraints of \cite{Perez2017} on the mass of the DM sterile neutrino (see \cref{fig:RAR_vs_P2017}). Such new bounds or relaxation in the (extended) $\nu$MSM parameters are subject entirely to both, the novel dense \textit{quantum} core at the center of the fermionic RAR+SIDM halo (see \cref{sec:bounds}), and to the interacting nature of the particles. The relaxed mass range consistently covers the one derived in \cite{Arguelles2016}, in which the degenerate fermion (ino) gas in the Galactic core provides an alternative to the central BH hypothesis. 


The intensity of the DM decay flux expected from an individual halo depends mainly on the DM density distribution in it. To discuss in detail the theoretical uncertainties in the calculation, we have estimated the signal by assuming different DM density profiles. Concretely, in \cref{sec:decay} we have described the signal with the decay width due to the Higgs portal interactions of the $\nu$MSM model, predicted the X-ray decay line, and presented different $S_{\rm DM}$ factor parametrizations. 

In \cref{subsec:vector} we have considered the RAR+SIDM profiles (generalising the traditional RAR ones \cite{Arguelles2016}), shown there to provide an excellent fit to the observed Milky Way rotation curve including the motion of the closest objects (S-stars) to SgrA*, without assuming the BH hypothesis (see \cref{fig:vrot-SIDM}). We further stressed that the novel \textit{dense quantum core - diluted halo} morphology present in the RAR+SIDM profiles, are different with respect to typical cored WDM halos (i.e. with a flattening in the inner density profiles) as arising from classical N-body simulations (see e.g. \cite{2019PrPNP.104....1B} for a recent review). The former, should be understood in terms of extensions of standard cosmological simulations when shifting from classical to quantum particles, and therefore both kind of WDM halos cannot be compared on equal footing. Indeed, similar core-halo profiles as the ones here studied for fermions, were already found for bosons in N-body simulations in the context of fuzzy DM (or quantum wave DM) as reported in \cite{2014PhRvL.113z1302S}. While in the case of FDM the central core (so called soliton) is supported against gravity by the \textit{quantum pressure} arising from the Heisenberg uncertainty principle (see \cite{2018PhRvD..97h3519M} for a recent discussion), in the case of fermions (i.e. RAR-cores) the corresponding \textit{degeneracy pressure} is provided by the Pauli exclusion principle \citep{2015MNRAS.451..622R}\footnote{
	It can be further shown that specific collisionless relaxation mechanisms, such as gravitational cooling and violent relaxation, can lead to each of these novel \textit{quantum core - diluted halo}  profiles respectively \cite{2018PhRvD..97h3519M}.
}.

In \cref{sec:interactingDM},  we considered a specific self-interacting model for sterile neutrino DM, including dark-sector massive vector exchanges among the sterile neutrinos, in the presence of a Higgs portal, thus extending appropriately the model of ref.~\cite{amrr}, in which a mixing of the sterile neutrino with the SM sector had been ignored. We discussed the effects of such self interactions on the sterile neutrino parameter-space bounds. We also studied, rather briefly though, the cosmological implications of such types of models in which the sterile neutrino plays the role of a cosmological DM. It is possible that one may circumvent the requirement for universe overclosure in such cases, and produce the required DM abundance via dark-sector vector boson decays to sterile neutrinos, which may in turn lead to a further diminishing of the lower bounds on the mixing angle $\theta$, as compared to  the standard-$\nu$MSM studies  
Although a complete analysis, including potential implications for BBN, is pending, nonetheless our results indicate a potentially significant increase of the available parameter space,  as compared to the traditional $\nu$MSM model~\cite{Perez2017}. This allows for larger sterile neutrino masses, above \SI{50}{\kilo\eV}, i.e. compatible with the range obtained in \cite{Arguelles2016} within the RAR scenario from the rotation-curve analysis, including the motion of the closest S-stars to SgrA*.

Our calculations for the $S_{\rm DM}$ factor, considering all the different profiles, are presented in \cref{sec:signal}. We conclude that the profile choice yields important differences in this factor, as expected. Since the RAR+SIDM profiles exhibit compact cores at small radius we obtain the maximum value for the $S_{\rm DM}$ factor with respect to the other profiles when the compact core region is included. However, we obtain lower values assuming the RAR+SIDM profile when such regions are not considered in the calculation. The dependence of the $S_{\rm DM}$ factor with the integration of the minimum value of the angle $\theta_{\rm min}$ forming with the GC direction, is shown in the \cref{fig:Grafico_comp_perfiles}.

Comparing these $S_{\rm DM}$ factors with the X-ray flux observations is necessary to obtain the null-detection bounds for these novel DM profiles. We estimate in \cref{sec:bounds} the photon flux from the GC region (at few \si{\parsec} scales from SgrA*) and reinterpret the 0-bounce photons analysis as obtained originally in \cite{Perez2017} (corresponding to $\sim \SI{E2}{\parsec}$ off the Galaxy center) for different DM profile assumptions. Assuming that no signal is observed (i.e. we assume a null-detection hypothesis), in \cref{eq:theta_upper_bound} we discuss the dependencies of the bound respect to $m_s$, $S_{\rm DM}$ and $F_{obs}$. For the GC region (inner pc), weaker (upper) limits in the mixing angle are obtained when the RAR+SIDM profile is considered with respect to other selected DM profiles, since the bound is inversely proportional to $S_{\rm DM}$, which is larger within the RAR central core as explained above. Thus, we remark once more that the bounds obtained here hinge entirely on the assumption of RAR+SIDM profiles, and within the AMRR-$\nu$MSM self-interacting paradigm \cite{amrr}.


\section*{Acknowledgements}
The work of C.R.A is supported by the National Research Council of Science and Technology (CONICET-Argentina).
C.R.A and A.M. further acknowledges the hospitality of the ICRANet Headquarters, where finalization of the present work was taking part.
The work of N.E.M. is partially supported by the
U.K. Science and Technology Facilities Council (STFC) via the Grant ST/L000258/1.
N.E.M. also acknowledges currently the hospitality of IFIC Valencia through a Scientific Associateship (\emph{Doctor  Vinculado}).
A.K. is supported by the Erasmus Mundus Joint Doctorate Program by Grants Number 2014--0707 from the agency EACEA of the European Commission.
The research of A.M. is supported by Fundação para  a  Ciência  e  a  Tecnologia  (FCT)  through  national funds  (UID/FIS/04434/2013),  by  FEDER  through COMPETE2020 (POCI-01-0145-FEDER-007672) and by FCT project  with reference PTDC/FISOUT/29048/2017. 

\begin{appendices}

\section{S factor}
\label{appendix:SDM}

An algorithm was developed to perform the $S_{\rm DM}$ factor integral, comprising of a solid angle integral and an integral along the line of sight, as seen in \cref{eq:S_factor}.

Each integral is performed as a Riemann sum: an integral is approximated as the sum of the function values on a grid, times the spacing between elements of such grid, as in:
\begin{align}
&\int_{a}^{b} f(x) \d x \approx \sum_{i=1}^{n}f(x_{i})\Delta x_{i} \\
&\text{where}\ \ x_{i}\in [a,b] \ ,\ \Delta x_{i}=\frac{x_{i+1}-x_{i-1}}{2} \nonumber
\end{align}
And the process is trivially extended for double and triple integrals. On the limit $n\rightarrow\infty$ both expressions are equivalent if $x_{i}$ are evenly spaced between a and b.

This kind of approximations yield greater errors in areas where $f$ changes rapidly and $X_{i}$ evaluation points are scarce. Thus, it is critical for the accuracy of these algorithms to make a good choice of evaluation points $X_{i}$. We will start by analyzing the solid angle integral, and how it is possible to optimize the evaluation points for the Riemann sum.

First, spherical symmetry of the halo density profile can be used to evaluate the dependence on the azimuthal angle $\Phi$:
\begin{equation}
\int_{\theta=0}^{\theta_{\rm max}}\int_{\Phi=0}^{2\pi}\bar{S}(\theta,\Phi)\sin{\theta}\d\theta \d\Phi=2\pi\int_{\theta=0}^{\theta_{\rm max}}\bar{S}(\theta) ~.
\end{equation}
Then only remains to choose a suitable choice of evaluation points for $\theta$. For circular shaped regions we chose logarithmically spaced points between $\theta \approx \SI{E-6}{\arcsec}$ and $\theta_{\rm max}$ ($\SI{40}{\arcsec}$).\footnote{
	The lower angle corresponds to the shortest relevant radius for RAR profiles.
} This allows us to have a greater definition around the profile inner regions, where density is expected to change rapidly. Logarithmic spacing was also used for `donut' shaped regions mentioned in \cref{sec:SA_GC}.

A more complex analysis is required for the integral along the line of sight:
\begin{equation}
\bar{S}(\theta,\Phi)=\int_{x=0}^{x=\infty}\rho(r(x,\theta,\Phi))\d x ~.
\end{equation}
Here we find the same challenge in evaluation pints: it is necessary to have tightly packed points around the $x$ values closer to the GC. But other numerical problems arise in the definition of r: 
\begin{equation}
r=\sqrt{r_{\rm GC}^{2}+x^{2}-2xr_{\rm GC}\cos{\theta}} ~.
\label{eq:r_one}
\end{equation}
Where $r_{\rm GC}$ is the distance between the Sum and the GC ($\approx \SI{8E3}{\parsec}$). If parameters are such that we can access the inner regions of the halo profile ($\approx \SI{E-7}{\parsec}$), then this result is to be acquired from the subtraction of two similar quantities up to $\SI{E-14}{\kilo\parsec^2}$: $(r_{\rm GC}^2+x^2)$ and $2xr_{\rm GC}\cos{\theta}$. This would have resulted in floating point precision errors for ordinary data storage types. Then, it was necessary to find an expression for $r(x,\theta)$ that remained accurate on such scales.

\begin{figure}
\centering
\includegraphics[width=0.8\hsize,clip]{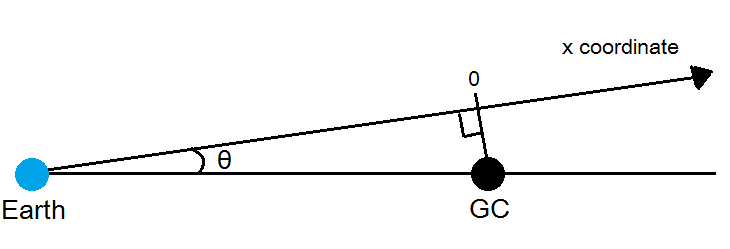}
\caption{Coordinate system election schematic. $x-\theta$ plane slice. X coordinate zero is set at the axis' closest point to the GC.}
\label{fig:Esquema_x_coord}
\end{figure}

So, we first redefine the zero of the x coordinate, so it is measured from the closest point to the GC, as shown in \cref{fig:Esquema_x_coord}.
Then, for small theta the expression in \cref{eq:r_one} can be approximated as:
\begin{align}
&r^{2}=r_{\rm GC}^{2} \left ( -2\epsilon(\theta) + \left( \frac{x}{r_{\rm GC}}\right )^2 - 4\frac{x}{r_{\rm GC}}\epsilon(\theta)\right )\nonumber\\
&\text{where}\  \epsilon(\theta)=\cos{\theta}-1=-\frac{\theta^2}{2}+\frac{\theta^4}{24}-...
\label{eq:r_2}
\end{align}
Where, for small $\theta$ and positive x it involves sums of positive expressions only.

Using this new definition of the x coordinate origin, we can solve the sampling problem using logarithmic spaced evaluation points around x=0. This kind of spacing was used on the intervals $X=[-r_{\rm GC},-10^{-7}]$ and $X=[10^{-7}, 2 r_{\rm halo}]$, with $r_{\rm halo}$ the MW DM halo radius ($\approx \SI{55}{\kilo\parsec}$). Then, Riemann sums were executed to evaluate the integral on these points using \cref{eq:r_2}.

\section{A sterile neutrino production mechanism: V-boson decay}
\label{Appx:Vboson_Decay}

The required Decay rate of a $V$-boson decaying into two sterile neutrinos, in view of the interaction term (\ref{vector_int}) is given by the standard formula:
\begin{equation}\label{width}
\Gamma = \frac{{\cal S}}{2E}\left[\int \prod_{i=1}^{m}\frac{\d^3p'_i}{2E_i' (2\pi)^3}\right]|\overline{{\cal M}}|^2 (2\pi)^4 \delta^{(4)}(\sum_{i=1}^m p_i' - p) ,
\end{equation}
where $p$ ($p_i^\prime, \, i=1, \dots m$) denote the four- momenta of the decaying particle (decay products),
${\cal S}$ is a statistical factor that equals $1/(n!)$ for each group of $n$ identical particles in the decay products, and $|\overline{{\cal M}}|^2$ is the square of the matrix element between initial and final states, averaged over initial-spin states and summed over final-spin ones.

For the evaluation of the amplitude ${\cal M}$ we use the Feynman rule for the vertex $V_\mu \, \overline{\nu_{1}}\,\nu_{1} $
\begin{equation}\label{rules}
- i g_V\, \gamma^\mu \, \Big( \frac{1 + \gamma_5}{2}  \Big) ~.
\end{equation}
To evaluate the decay width (\ref{width}) we use the following Casimir identities (in the following expressions, $u_a(p)$ ($v_b(p)$) denote polarization spinors (antispinors) appearing
in the solution of the free Dirac (or Majorana) equations):
\begin{align} \label{rules2}
&\sum_{i_A,i_B=\uparrow,\downarrow} \left({\overline u}^{(i_A)}(p_A)\Gamma_I u^{(i_B)}(p_B)\right)^\dagger \left({\overline u}^{(i_A)}(p_A)\Gamma_{II} u^{(i_B)}(p_B)\right)
\nonumber \\
&= {\rm Tr}\left[{\overline \Gamma}_I \left(\gamma^\mu {p_A}_\mu  + m_A\right)\Gamma_{II} \left(\gamma^\mu {p_B}_\mu + m_B\right)\right], \nonumber \\
& \sum_{i_A,i_B=\uparrow,\downarrow} \left({\overline v}^{(i_A)}(p_A)\Gamma_I u^{(i_B)}(p_B)\right)^\dagger \left({\overline v}^{(i_A)}(p_A)\Gamma_{II} u^{(i_B)}(p_B)\right) \nonumber \\
&= {\rm Tr}\left[{\overline \Gamma}_I \left(\gamma^\mu {p_A}_\mu  - m_A\right)\Gamma_{II} \left(\gamma^\mu {p_B}_\mu + m_B\right)\right] , \nonumber \\
& \sum_{i_A,i_B=\uparrow,\downarrow} \left({\overline v}^{(i_A)}(p_A)\Gamma_I v^{(i_B)}(p_B)\right)^\dagger \left({\overline v}^{(i_A)}(p_A)\Gamma_{II} v^{(i_B)}(p_B)\right) \nonumber \\
& = {\rm Tr}\left[{\overline \Gamma}_I \left(\gamma^\mu {p_A}_\mu  - m_A\right)\Gamma_{II} \left(\gamma^\mu {p_B}_\mu - m_B\right)\right],
\end{align} for any two $4 \times 4$ matrices $\Gamma_I, \Gamma_{II}$, where
$\overline{\Gamma}_I \equiv \gamma^0 \Gamma_I^\dagger \gamma^0$, with $\gamma^\mu$ the $4\times 4$ Dirac $\gamma$-matrices.
For our purposes it suffices to calculate the width (\ref{width}) in flat Minkowski space time. In this case, we have the properties
\begin{align}
	\{\gamma^\mu, \, \gamma^\nu \} &= 2\eta^{\mu\nu}\textbf{1}\\
	\{\gamma^\mu, \, \gamma^5 \} &= 0\\
	{\overline \gamma}^\mu \equiv \gamma^0 (\gamma^\mu)^\dagger \gamma^0 &= \gamma^\mu\\
	\overline{\gamma^\mu \gamma^5} \equiv \gamma^0 (\gamma^\mu \gamma^5)^\dagger \gamma^0 &= \gamma^\mu \gamma^5\\
	{\rm Tr}[{\rm \textbf{1}}] &= 4\\
	{\rm Tr}\left(\gamma^\mu \,\gamma^\nu \right) &= 4\eta^{\mu\nu}\\
	{\rm Tr}\left(\gamma^\mu \gamma^\lambda \gamma^\nu \gamma^\rho \right) &= 4\left(\eta^{\mu\lambda}\eta^{\nu\rho} - \eta^{\mu\nu}\eta^{\lambda\rho} + \eta^{\mu\rho}\eta^{\nu\lambda}\right)
\end{align}
where $\gamma^5 = i \gamma^0 \, \gamma^1 \, \gamma^2 \, \gamma^3 $, the trace ${\rm Tr}$ is over spinor indices, \textbf{1} denotes the $4 \times 4$ identity  matrix in spinor space and $\eta^{\mu\nu} $ is the Minkowski metric with the signature convention $(\,+, \,-, \,-, \,-)$.
Note that the trace ${\rm Tr} \left( \gamma^\mu \gamma^\nu \gamma^5\right) = 0$ while ${\rm Tr} \left( \gamma^\mu\gamma^\lambda \gamma^\nu \gamma^\rho \gamma^5\right)$ is totally antisymmetric in the Lorentz indices.

Since the phenomenological considerations of \cite{amrr} require the vector boson mass $m_V$ to be much larger (at least four orders of magnitude)  than the sterile neutrino DM mass $m$,  $m_V \gg m $, we may treat the fermionic product of the decay $V_\mu \, \overline{\nu_{1}}\,  \nu_{1} $ as \emph{practically massless}. Hence, applying the identities of \cref{rules2} in this case,
yields: 
\begin{equation}
	- i {\cal M} =  \epsilon_V^\mu (p) \, \bar{u}_{f_1}^{(i1)} (p_1^\prime)  \, \left( -i g_V\, \gamma_\mu \, \right)  \, \left(\frac{1 + \gamma^5}{2}\right)   v_{f_2}^{(i2)} (p^\prime_2)\,,
\end{equation} 
where $\epsilon^\mu_V $ is the polarisation of the massive $V$-boson. and $f_{1,2}$ denote the fermionic decay products in the three processes. The fermions $f_i$ are all massless.
The square of the initial-spin-averaged and final-spin-summed amplitude entering \cref{width} reads:
 \begin{equation}\label{avampl}
   |\overline{{\cal M}}|^2 = \frac{1}{3} \sum_{i_1, i_2 = \uparrow, \downarrow} |{\cal M}|^2,
   \end{equation}
where the factor $1/3$ is due to the fact that we have $2 s + 1$ (with $s=1$) initial spins of the massive vector boson to average over. Taking into account  the identities of \cref{rules2}, with the matrices $\Gamma_I, \Gamma_{II}$ being $\gamma^\mu, \gamma^\nu$ and $\gamma^\mu \gamma^5, \gamma^\nu \gamma^5$,
as well as the fact that the sum over vector-boson-$V$ polarisation (spin) states is 
\begin{equation*}
	\sum_{\rm spin} \epsilon_\mu (p) \, \epsilon_\nu (p) =  -\eta_{\mu\nu} + \frac{p_\mu p_\nu}{m_V^2} ,
\end{equation*}
we may evaluate the amplitude (\ref{avampl}) as (from now on we omit the particle-species index from the polarisation tensors of spinors for simplicity)
\begin{align*}
	|\overline{{\cal M}}|^2 
	&= \frac{g_V^2}{3} \, \sum_{i_1, i_2 = \uparrow, \downarrow} \left[ \bar{u}^{(i1)} (p_1^\prime) \gamma_\mu \frac{1 +\gamma^5}{2} v^{(i2)} (p_2^\prime) \right]^\dagger \\
	&\phantom{=} \left[ \bar{u}^{(i1)} (p_1^\prime) \gamma_\nu \frac{1+ \gamma^5}{2} v^{(i2)} (p_2^\prime) \right]  \epsilon_V^\mu (p) \epsilon_V^\nu (p)\\
	&=\frac{g_V^2 }{3}  {\rm Tr} \, \Big[\gamma_\mu \frac{1 + \gamma^5}{2} (\gamma_\rho p_2^{\prime \, \rho}  ) \, \gamma_\nu \,
 \frac{1+ \gamma^5}{2} \, \gamma_\sigma \, p_1^{\prime \, \sigma} \Big]\\
	&\phantom{=} \times  \left( - \eta^{\mu \nu} + \frac{p_V^\mu p_V^\nu}{m_V^2} \right)\\
	&= \frac{g_V^2}{6} {\rm Tr} [\gamma_\mu (1+ \gamma^5) \, \gamma_\rho \gamma_\nu \gamma_\sigma  ] \ p_2^{\prime \, \rho} p_1^{\prime \, \sigma}  \\
	&\phantom{=} \times \left( - \eta^{\mu \nu} + \frac{p_V^\mu p_V^\nu}{m_V^2} \right)\, \\
	&= \frac{g_V^2}{6} {\rm Tr} [\gamma_\mu \, \gamma_\rho \gamma_\nu \gamma_\sigma  ] \ p_2^{\prime \, \rho} p_1^{\prime \, \sigma} \,  \left( - \eta^{\mu \nu} + \frac{p_V^\mu p_V^\nu}{m_V^2} \right)\, .
\end{align*} 
In the last simplification, we used the anti-commutation properties of $\gamma^5$ with $\gamma_\mu$, and the fact that $\left( \frac{1 + \gamma^5}{2} \right)^2 = \frac{1 + \gamma^5}{2} $ and $\frac{1+\gamma^5}{2} \frac{1-\gamma^5}{2} =0$.
Above we also took into account that the trace containing $\gamma^5$ is zero because it gives rise to a totally antisymmetric tensor (rubric) which is contracted with a symmetric tensor with respect to the $\mu, \nu$, indices,  $\left( - \eta^{\mu \nu} + \frac{p_V^\mu p_V^\nu}{m_V^2} \right)$.
Using the identities of Dirac matrices, given previously, we then obtain
\begin{align}\label{finalampl}
	|\overline{{\cal M}}|^2
	&= \frac{2\,g_V^2}{3} (p_{2 \mu}^\prime p_{1 \nu}^\prime + p_{2 \nu}^\prime  p_{1 \mu}^\prime  - \eta_{\mu \nu} p_1 ^\prime \cdot p_2^\prime  ) \\
	&\phantom{=} \times \left( - \eta^{\mu \nu} + \frac{p^\mu p^\nu}{m_V^2} \right) \nonumber\\
	&= \frac{2\,g_V^2}{3}\left( p_2^\prime \cdot p_1^\prime  + 2 \frac{p_2^\prime  \cdot p \, p_1^\prime \cdot p}{m_V^2} \right) \nonumber
 \end{align}
where we used the on-shell condition for the $V$ momentum $ p^\mu \eta_{\mu\nu} p^\nu = m_V^2$.
Using the conservation of energy momentum in the vertex ($p$ incoming, $p_1^\prime, p_2^\prime$ outgoing),
\begin{equation}\label{cons}
 p^\mu - p_1^{\prime \, \mu} - p_2^{\prime \, \mu} = 0  ,
 \end{equation}
we square it (in a covariant way) to derive:
\begin{align*}
 0 	&=  p^\prime_2 \cdot p^\prime_2 + p \cdot p  +  p^\prime_1 \cdot p^\prime_1 + 2 p^\prime_2 \cdot p^\prime_1 - 2 p  \cdot (p^\prime_2 +  p^\prime_1) \\
		&= m_V^2  + 2 p^\prime_2 \cdot p^\prime_1 - 2 p \cdot p =  - m_V^2 + 2 p^\prime_2 \cdot p^\prime_1
\end{align*}\begin{equation*}
	\Rightarrow p^\prime_2 \cdot p^\prime_1 = m_V^2 /2
\end{equation*} where we used the on-shell conditions (in our conventions for the metric $(+1,-1,-1,-1)$) \begin{equation*}
	p \cdot p = m_V^2 , \quad p^\prime_1 \cdot p^\prime_1 = 0, \quad p^\prime_2 \cdot p^\prime_2 = 0 .
\end{equation*}
In a similar way by writing the square as 
\begin{align*}
	0	&=  p^\prime_2 \cdot p^\prime_2 + p \cdot p +  p^\prime_1 \cdot p^\prime_1 - 2 p_1^\prime  \cdot (p - p^\prime_2 )  - 2 p \cdot p^\prime_2 \\
		&= p^\prime_2 \cdot p^\prime_2 + p \cdot p  - p^\prime_1 \cdot p^\prime_1 - 2 p \cdot p^\prime_2 \\
		&\Rightarrow  p \cdot p^\prime_2 =  m_V^2 /2 .
\end{align*} And finally, by writing the square as: \begin{align*}
	0 &= p^\prime_2 \cdot p^\prime_2 + p \cdot p +  p^\prime_1 \cdot p^\prime_1 - 2  p^\prime_2 \cdot  (p - p^\prime_1 ) - 2 p^\prime_1 \cdot  p \\ 
		&= m_V^2 - 2 p^\prime_2 \cdot p^\prime_2  - 2 p \cdot p^\prime_1 \, \\
		&\Rightarrow \, p^\prime_1 \cdot p =  m_V^2 /2~.
\end{align*} 
Then, the  amplitude can be written in terms of masses
\begin{equation}
	\label{finalwidth}
	|\overline{{\cal M}}|^2  = \frac{2\,g_V^2}{3}\, m_V^2\, .
\end{equation}

In the rest frame of the $V$-boson ($E_V = m_V$,  $\vec{p}_V = 0$)
the  phase space integration in \cref{width} is done by first performing the spatial
delta function integration $$\int \d^3 p^\prime_{2}\, \delta ^{(3)} (  - p_1^\prime -  p_2^\prime )~, $$ which simply implies
that the spatial momenta of the decay products (which are massless particles) are equal in magnitude  $|\vec{p}_1^{\prime}| = |\vec{p}_2^{\prime}|$.

In the case of Majorana sterile neutrinos, there is one group of two ($n=2$) identical particles in the products of the $V$-vector-boson decay so the
statistical factor $\mathcal S$ in the definition of the width (\ref{width}) is
$\mathcal S=\frac{1}{2}$ (for Dirac type ``inos'' $\mathcal S=1$, as in that case there are no identical particles among the decay products). Treating the neutrino as practically massless inside the phase-space integration, which suffices for our approximate discussion here,
we then obtain: 
\begin{align}
	\Gamma 	&= \frac{1}{16 \, m_V} \int 4\pi |\vec p_2'|^2 \d|\vec p_2'|  \delta (m_V - 2|\vec{p_2'}|) \nonumber \\
	&\phantom{=} \times \frac{1}{4\pi^2\, |\vec{p_2'}|^2} |\overline{{\cal M}}|^2 \nonumber \\
	\label{result}
					&= \frac{1}{32 \, \pi \, m_V} \, |\overline{{\cal M}}|^2 \simeq \frac{g_V^2} {48\, \pi}\, m_V
\end{align} 
where we used that $ \int \d|\vec p_2^\prime|   \delta (m_V - 2|\vec{p_2^\prime}|) = \frac{1}{2}$.

\end{appendices}

\bibliographystyle{model1-num-names}
\bibliography{biblio,Report2,Report3,rev2}

\end{document}